%Paper: hep-th/9403197
%From: "Adel Bilal" <bilal@puhep1.Princeton.EDU>
%Date: Thu, 31 Mar 94 13:22:06 -0500
%Date (revised): Tue, 19 Apr 94 14:03:11 -0400

%%%%%%%%%%%%%%%%%%%%%%%%%%%%%%%%%%%%%%%%%%%%%%%%%%%%%%%%%%%%%%%

\input phyzzx

%%%%%%%%%%%%%%%%%%%%%%%%%%%%%%%%%%%%%%%%%%%%%%%%%%%%%%%%%%%%%%%%
% This will make your PHYZZX pagesize wider and longer
% It is OPTIONAL. It redefines the papers macro
%
\catcode`\@=11 % This allows us to modify PLAIN macros.
\def\papers{\papersize\headline=\paperheadline\footline=\paperfootline}
\def\papersize{\hsize=40pc \vsize=53pc \hoffset=0pc \voffset=1pc
   \advance\hoffset by\HOFFSET \advance\voffset by\VOFFSET
   \pagebottomfiller=0pc
   \skip\footins=\bigskipamount \normalspace }
\catcode`\@=12 % at signs are no longer letters
\papers
%%%%%%%%%%%%%%%%%%%%%%%%%%%%%%%%%%%%%%%%%%%%%%%%%%%%%%%%%%%%%%%%

\def\to{\rightarrow}

\vsize=21.5cm
\hsize=15.cm

\tolerance=500000
\overfullrule=0pt

\Pubnum={PUPT-1452 \cr
hep-th@xxx/9403197 \cr
March 1994}

\date={}
\pubtype={}
\titlepage
\title{NON-LOCAL MATRIX GENERALIZATIONS OF $W$-ALGEBRAS}
\author{{
Adel~Bilal}\foot{
 on leave of absence from
Laboratoire de Physique Th\'eorique de l'Ecole
Normale Sup\'erieure, \nextline 24 rue Lhomond, 75231
Paris Cedex 05, France
(unit\'e propre du CNRS)\nextline
e-mail: bilal@puhep1.princeton.edu
}}
\address{\it Joseph Henry Laboratories\break
Princeton University\break
Princeton, NJ 08544, USA}

\vskip 3.mm
\abstract{There is a standard way to define two symplectic
(hamiltonian) structures, the first and second Gelfand-Dikii brackets,
on the space of ordinary $m^{\rm th}$-order linear differential operators
$L = -d^m + U_1 d^{m-1} + U_2 d^{m-2} + \ldots + U_m$. In this paper, I
consider in detail the case where the $U_k$ are $n\times
n$-matrix-valued functions, with particular emphasis on the (more
interesting) second Gelfand-Dikii bracket. Of particular interest is the
reduction to the symplectic submanifold $U_1=0$. This reduction gives
rise to matrix generalizations of (the classical version of) the {\it
non-linear} $W_m$-algebras, called $V_{m,n}$-algebras. The
non-commutativity of the matrices leads to {\it non-local} terms in
these $V_{m,n}$-algebras. I show that these algebras contain a conformal
Virasoro subalgebra and that combinations $W_k$ of the $U_k$ can be
formed that are  $n\times n$-matrices of conformally primary fields of
spin $k$, in analogy with the scalar case $n=1$. In general however,
the $V_{m,n}$-algebras have a much richer structure than the
$W_m$-algebras as can be seen on the examples of the {\it non-linear} and
{\it non-local} Poisson brackets $\{(U_2)_{ab}(\sigma) ,
(U_2)_{cd}(\sigma')\}\ $, $\ \{(U_2)_{ab}(\sigma) ,
(W_3)_{cd}(\sigma')\}$ and $\{(W_3)_{ab}(\sigma) ,
(W_3)_{cd}(\sigma')\}$ which I work out explicitly for all $m$ and $n$. A
matrix Miura transformation is derived, mapping these complicated
(second Gelfand-Dikii) brackets of the $U_k$ to a set of
much simpler Poisson brackets, providing the
analogue of the free-field representation of the $W_m$-algebras.}

\endpage
\pagenumber=1

 \def\PL #1 #2 #3 {Phys.~Lett.~{\bf #1} (#2) #3}
 \def\NP #1 #2 #3 {Nucl.~Phys.~{\bf #1} (#2) #3}
 \def\PR #1 #2 #3 {Phys.~Rev.~{\bf #1} (#2) #3}
 \def\PRL #1 #2 #3 {Phys.~Rev.~Lett.~{\bf #1} (#2) #3}
 \def\CMP #1 #2 #3 {Comm.~Math.~Phys.~{\bf #1} (#2) #3}
 \def\IJMP #1 #2 #3 {Int.~J.~Mod.~Phys.~{\bf #1} (#2) #3}
 \def\JETP #1 #2 #3 {Sov.~Phys.~JETP.~{\bf #1} (#2) #3}
 \def\PRS #1 #2 #3 {Proc.~Roy.~Soc.~{\bf #1} (#2) #3}
 \def\IM #1 #2 #3 {Inv.~Math.~{\bf #1} (#2) #3}
 \def\JFA #1 #2 #3 {J.~Funkt.~Anal.~{\bf #1} (#2) #3}
 \def\FAA #1 #2 #3 {Funct.~Anal.~Applic.~{\bf #1} (#2) #3}
 \def\LMP #1 #2 #3 {Lett.~Math.~Phys.~{\bf #1} (#2) #3}
 \def\IJMP #1 #2 #3 {Int.~J.~Mod.~Phys.~{\bf #1} (#2) #3}
 \def\FAA #1 #2 #3 {Funct.~Anal.~Appl.~{\bf #1} (#2) #3}
 \def\AP #1 #2 #3 {Ann.~Phys.~{\bf #1} (#2) #3}
 \def\MPL #1 #2 #3 {Mod.~Phys.~Lett.~{\bf #1} (#2) #3}

\def\d{\partial}

\def\s{\sigma}
\def\sp{\sigma'}
\def\l{\lambda}
\def\is{\int {\rm d} \sigma\, }

\def\ds{\delta(\s-\s')}
\def\dsp{\delta'(\s-\s')}

\def\dsppp{\delta'''(\s-\s')}
\def\es{\epsilon(\s-\s')}
\def\e{\epsilon}

\def\gd{\gamma^2}
\def\gmd{\gamma^{-2}}
\def\rmd{{\rm d}}
\def\rd{\sqrt{2}}
\def\vp{{V^+}}
\def\vm{{V^-}}
\def\GD{Gelfand-Dikii\ }
\def\P{\Psi}
\def\dd #1 #2{{\delta #1\over \delta #2}}
\def\bin #1 #2{{ #1\choose #2}}
\def\tr{{\rm tr}\ }
\def\res{{\rm res}\ }

\def\id{{\bf 1}}
\def\Lemma #1  {\noindent{\bf Lemma #1 :\ }}
\def\Proof{\noindent{\it Proof :\ }}
\def\v{{\cal V}(f)}
\def\vt{{\tilde{\cal V}}(f)}
\def\ft{{\tilde f } }
\def\gt{{\tilde g } }
\def\P{{\cal P} }
\def\dh{{\hat \d} }

{\chapter{Introduction}}

Since their discovery by Zamolodchikov
\REF\ZAM{A.B. Zamolodchikov, {\it Infinite additional symmetries
in two-dimensional conformal field theory}, Theor. Math. Phys. {\bf 65}
(1985) 1205. }
[\ZAM], $W$-algebras have been an active field of
investigation in theoretical and mathematical physics (see refs.
\REF\TRIESTE{A. Bilal, {\it Introduction to $W$-algebras}, in: Proc.
Trieste Spring School on String Theory and Quantum Gravity, April 1991,
J. Harvey et al eds., World Scientific, p.245-280.}
\REF\BS{P. Bouwknegt
and K. Schoutens, {\it $W$-symmetry in conformal field theory}, Phys.
Rep. {\bf 223} (1993) 183.} [\BS,\TRIESTE] for reviews). They are
extensions of the conformal Virasoro algebra by higher spin fields
$W_k$. The commutator of two such higher spin fields is a {\it local}
expression involving {\it non-linear} differential polynomials of the
$W_l$. $W$-algebras were found to arise naturally in the context of the
$1+1$-dimensional Toda field theories
\REF\BGTODA{A. Bilal and J.-L. Gervais, {\it Systematic approach to
conformal systems with extended Virasoro symmetries}, Phys. Lett.
{\bf B206} (1988) 412; {\it Systematic construction of $c=\infty$
conformal systems from classical Toda field theories},  Nucl. Phys.
{\bf B314} (1989) 646; {\it Systematic construction of conformal
theories with higher-spin Virasoro symmetries}, Nucl. Phys. {\bf B318}
(1989) 579.}
[\BGTODA] where the higher spin fields
appeared  as coefficients of a linear differential operator
$L=-\d^m+\sum_{k=2}^m u_k \d^{m-k}$ that annihilates  the first Toda
field $e^{-\phi_1}$. At a more formal level, the classical (i.e. Poisson
bracket) version of $W$-algebras were shown
\REF\BAK{I. Bakas, {\it Higher spin fields and the \GD algebras},
Commun. Math. Phys. {\bf 123} (1989) 627.}
\REF\MAT{P. Mathieu, {\it Extended classical conformal algebras and
the second hamiltonian structure of Lax equations}, Phys. Lett. {\bf
B208} (1988) 101.}
\REF\DIZ{P. Di Francesco, C. Itzykson and J.-B.
Zuber, {\it Classical $W$-algebras}, Commun. Math. Phys.
{\bf 140} (1991) 543 .}
[\BAK,\MAT,\DIZ] to be given by the second \GD bracket associated with
the linear differential operator $L$. This also implied close
connections with the generalized KdV hierarchies \REF\SW{G. Segal and G.
Wilson, {\it Loop groups and equations of KdV type}, Inst. Hautes Etudes
Sci. Publ. Math. {\bf 61} (1985) 5.} [\SW].

Recently, in the study of the simplest $1+1$-dimensional {\it
non-abelian} Toda field theory
\REF\NAT{A. Bilal, {\it Non-abelian Toda theory: a completely
integrable model for strings on a black hole background}, Princeton
University preprint PUPT-1434 (December 1993), hep-th@xxx/9312108, to
appear in Nucl. Phys. B.} [\NAT], a related but more general structure
was discovered. This non-abelian Toda theory has 3 conserved left-moving
currents $T,V^+, V^-$ and 3 conserved right-moving ones $\bar T, \bar
V^+, \bar V^-$. The study of ref. \NAT\ was purely at the classical level
(Poisson brackets), and it was found that $T, V^+$ and $V^-$ form a {\it
non-linear} and {\it non-local} Poisson bracket algebra\foot{
The $\bar T, \bar V^+, \bar V^-$ form an isomorphic algebra, while
the Poisson bracket of a left-moving with a right-moving current
vanishes.}:
$$\eqalign{
\gmd \{T(\s)\, ,\, T(\s')\}  &=
(\d_\s-\d_{\s'})\left[ T(\s') \ds\right]-{1\over 2} \dsppp \cr
\gmd \{T(\s)\, ,\, V^\pm(\s')\}  &=
(\d_\s-\d_{\s'})\left[ V^\pm(\s') \ds\right]\cr
\gmd \{V^\pm(\s)\, ,\, V^\pm(\s')\}&=\es
V^\pm(\s)V^\pm(\s')\cr
\gmd \{V^\pm(\s)\, ,\, V^\mp(\s')\}&=-\es
V^\pm(\s)V^\mp(\s')\cr
&\phantom{=}+(\d_\s-\d_{\s'})\left[ T(\s') \ds\right] -{1\over 2}
\dsppp \ .\cr }
\eqn\ui$$
One sees that $T$ generates the conformal algebra. More precisely, if
$\s$ takes values on the unit circle $S^1$, then
$L_r=\gmd \int_{-\pi}^\pi \rmd\s [T(\s)+{1\over 4}]e^{ir\s}$
generates a Poisson bracket Virasoro algebra $i\{L_r,L_s\}=
(r-s)L_{r+s} +{c\over 12} (r^3-r)\delta_{r+s,0}$ with central charge
$c=12\pi \gmd$ where $\gd$ is a
(classically arbitrary) scale factor. The bracket \ui\ also shows
that $\vp$ and $\vm$ are spin-2 conformal primary fields. The Poisson
brackets of $\vp$ and $\vm$ with themselves contain the new non-local
terms involving $\es\sim 2\d^{-1}\ds$.\foot{The precise definition of
$\es$ depends on the boundary conditions. For example, on the space
of functions on $S^1$ without constant Fourier mode, $\d^{-1}$ is
well-defined and $\es={1\over i\pi}\sum_{n\ne 0}{1\over
n}e^{in(\s-\sp)}$.} The complete mode expansion of the algebra \ui\ was
written in ref. \NAT\ (eq. (3.24)). To emphasize the similarities
(non-linearity) and differences (non-locality) with the $W$-algebras,
this algebra was called $V$-algebra [\NAT].

It was conjectured in ref. \NAT\ and confirmed in ref.
\REF\MCK{A. Bilal, {\it Multi-component KdV hierarchy, $V$-algebra and
non-abelian Toda theory}, Princeton University preprint PUPT-1446
(January 1994), hep-th@xxx/9401167.}
\MCK\ that the $V$-algebra \ui\ is again associated with a linear
differential operator $L=-\d^2+U$ (where $\d\equiv\d_\s=\d/\d\s$), but
this time $U$ being a matrix, namely the $2\times 2$ matrix
$U=\pmatrix{ T& -\rd V^+\cr -\rd V^- & T \cr }$. The non-locality of
the algebra turned out to be related to the non-commutativity of
matrices. More precisely, there is a standard way
\REF\GELDA{I.M. Gel'fand and L.A. Dikii, {\it Fractional powers of
operators and hamiltonian systems},
 Funct. Anal. Applic. {\bf 10} (1976) 259.}
\REF\GELF{I.M. Gel'fand and L.A. Dikii, {\it The resolvent and
hamiltonian systems}, Funct. Anal. Applic. {\bf 11} (1977) 93.}
\REF\ADLER{M. Adler, {\it On a trace functional for
pseudo-differential operators and the symplectic structure of the
Korteweg-de Vries type equations}, Invent. math. {\bf 50} (1979)
219.}
\REF\LEBMAN{D.R. Lebedev and Yu.I. Manin, {\it Gel'fand-Dikii
hamiltonian operator and the coadjoint representation of the Volterra
group}, Funct. Anal. Applic. {\bf 13} (1979) 268.}
\REF\GELDB{I.M. Gel'fand and L.A. Dikii, {\it A family of
hamiltonian structures connected with integrable non-linear
differential equations}, Inst. Appl. Math. Acad. Sci. USSR preprint
n$^{\rm o}$ 136, 1978 (in Russian).}
 \REF\KW{B.A. Kupershmidt and G. Wilson, {\it Modifying Lax equations
and the second hamiltonian structure}, Invent. Math. {\bf 62} (1981)
403.}
\REF\DIK{L.A. Dikii, {\it A short proof of a Kupershmidt-Wilson
theorem}, Commun. Math. Phys. {\bf 87} (1982) 127.}
[\GELDA-\DIK] to associate
 two symplectic structures (Poisson brackets), the first and second
\GD bracket, to any linear differential operator with scalar
coefficients $u_k$. It was shown in ref. \NAT\ that the analogous
construction of the second \GD bracket for $L=-\d^2+U$, with $U$ the
above $2\times 2$-matrix, precisely is the $V$-algebra \ui. Using the
$2\times 2$-matrix $U$ written above, the algebra \ui\ can be written
more compactly. Since $U$ is constrained by $\tr \s_3 U=0$, it is
convenient to introduce $2\times 2$-matrix-valued (test-) functions $F$
and $G$ subject to the same constraint: $\tr \s_3 F=\tr \s_3 G=0$. Then
$$\eqalign{
\gmd\{\int\tr FU, \int\tr GU\}=\int\tr\Big(
&GF'''+[F,U]\d^{-1}[G,U]\cr
&-(F'G-FG'+GF'-G'F)U\Big)\ .\cr}
\eqn\uii$$
where the commutators $[F,U]$ and $[G,U]$ on the r.h.s. are simply the
commutators of the $2\times 2$-matrices.

 It is the purpose of this paper
to generalize these developments to a linear differential operator
$L=-\d^m+\sum_{k=1}^mU_k\d^{m-k}$ of arbitrary order $m\ge 1$ with
coefficients $U_k$ that are $n\times n$-matrices. The corresponding
algebras will be called $V_{n,m}$-algebras.\foot{More precisely, I
reserve the name $V_{n,m}$-algebra for the reduction to the submanifold
where $U_1=0$, see below.}
Symplectic structures (Poisson brackets) associated with $n\times
n$-matrix $m^{\rm th}$-order differential operators have beeen
studied in the mathematical literature by Gelfand and Dikii
[\GELF]. The Poisson bracket they define is now called the first \GD
bracket, and leads to a {\it linear} algebra. In their paper [\GELF]
they also give the asymptotic expansion of the resolvent of $L$ and a
recursion relation for an infinite sequence of hamiltonians in
involution with respect to this first \GD bracket. Here, I am mainly
interested in the second \GD bracket [\GELDB], and to the best of my
knowledge this bracket has never been worked out so far for the matrix
case. For $m=2$ and $L=-\d^2+U$, it was shown in ref. \MCK\ that the
models are bihamiltonian, and that the second \GD bracket actually
follows from the recursion relations for the resolvent, or those for the
hamiltonians.\foot{Earlier related studies for $m=2$ can be found in
\REF\OMG{E. Olmedilla, L. Martinez Alonso and F. Guil, {\it
Infinite-dimensional hamiltonian systems associated with matrix
Schroedinger operators}, Nuovo Cim. {\bf 61B} (1981) 49.}
\REF\BR{M. Bruschi and O. Ragnisco, {\it The hamiltonian structure of
the nonabelian Toda hierarchy}, J. Math Phys. {\bf 24} (1983) 1414.}
\REF\CD{F. Calogero and A. Degasperis, {\it Nonlinear evolution
equations solvable by the inverse spectral transform associated with the
multichannel Schroedinger problem, and properties of their solution},
Lett. Nuovo Cim. {\bf 15} (1976) 65; {\it Nonlinear evolution equations
solvable by the inverse spectral transform - II}, Nuovo Cim. {\bf 39B}
(1977) 1.}
 refs. \OMG, \BR\ and \CD. See also the Note Added.}
 For $m\ge 3$, these recursion relations are much more
complex [\GELF], and it is not clear whether the second \GD bracket
given in the present paper could also be extracted from the formulas
of ref. \GELF. In any case, the construction given here is a
straightforward generalization of the scalar case. Let me remark that
all the developments of the present paper can be generalized by
replacing the $n\times n$-matrices $U_k$ by operators $\hat U_k$
acting in some Hilbert space, provided the products of these
operators and their traces, as well as the functional derivatives
$\delta/\delta \hat U_k$ are well-defined.

Let me note that for $m=2$, it was shown in refs \MCK, \OMG, \CD\
how to construct  an infinite sequence of hamiltonians with
respect to both \GD brackets. This let to matrix KdV hierarchies. The
present developments are connected with matrix versions of the
generalized KdV  hierarchies (i.e. matrix \GD hierarchies)  or matrix KP
hierarchies. The latter were recently studied from a somewhat different
point of view by Kac and van de Leur \REF\KVL{V.G. Kac and J.W. van de
Leur, {\it The $n$-component KP hierarchy and representation theory},
MIT preprint (August 1993), hep-th@xxx/9308137.}
[\KVL], and the present paper is quite complementary to theirs.

In section 2, I will work out the second \GD bracket\foot{The first
\GD bracket will also be given but it seems to be less interesting.},
first for generic $U_1, \ldots U_m$ and then for differential
operators $L$ with $U_1=0$. The reduction from the general case to
the case $U_1=0$ is non-trivial, as usual. In particular, it
introduces the integral ($\d^{-1}$) of a commutator of two matrices.
It is thus the reduction to the symplectic submanifold $U_1=0$ together
with the non-commutativity of matrices that leads to the non-localities
in the second \GD bracket. I note that for $m=1$ (and $U_1\ne 0$) one
simply recovers a $gl(n)$ Kac-Moody algebra (cf. ref.
\REF\DS{V.G. Drinfel'd and V.V. Sokolov, {\it Lie algebras and
equations of Korteweg-de Vries type}, Sov. Math. {\bf 30} (1985) 1975.}
\DS).

In section 3, I prove that a Miura-type transformation
$L=-(\d-P_1)(\d-P_2)\ldots (\d-P_m)$ maps the (relatively)
complicated symplectic structure given by the Poisson brackets of the
$U_k$ to a much simpler symplectic structure given by the Poisson
brackets of a set of $m$ decoupled fields $P_i$, each $P_i$ being a
$n\times n$-matrix. This provides the analogue of the usual free
field realization: whereas in the scalar case ($n=1$) the $P_i$ are
just $m$ free fields, i.e. $m$ collections of harmonic oscillators,
or in other words they form $m$ copies of a (Poisson bracket) $U(1)$
current algebra, here, due to the matrix character, the $P_i$ form
$m$ copies of a (Poisson bracket) $gl(n)$ current algebra. I also
discuss the reduction to $\sum_i P_i=0$ corresponding to $U_1=0$.
Since the Jacobi identity is obviously satisfied by the Poisson
bracket of the $P_i$, as an important corollary, the Miura
transformation immediately implies the Jacobi identity for the second
\GD bracket of the $U_k$, which was not obvious a priori.

In section 4, I will discuss the conformal properties. It turns out
that (for the case $U_1=0$)
 $T\sim \tr U_2$ generates the conformal (Virasoro) algebra, and
the Poisson brackets of $\tr U_2$ with any matrix element of any $U_k$
is relatively easy to spell out. They turn out to be exactly the same as
in the scalar case, and I conclude that appropriately symmetrized
combinations $W_k$ can be formed so that all of their matrix elements
are conformal primary fields of weight (spin) $k$.
This analogy with the scalar case is due to the fact that the
conformal properties are determined by $T\sim \tr U_2\equiv \tr \id
U_2$ and the unit matrix $\id$ always commutes.

In section 5, I spell out the Poisson bracket algebra of the matrix
elements of $U_2$ and $U_3$ for any $m\ge 2$ (with the restriction
$U_1=0$ which is the more interesting case). Again, this is done more
compactly by considering $\int\tr FU_2$ and $\int\tr FU_3$. For $m=3$,
and in the primary basis $U_2, W_3$, the result reads ($a$ is related
to the scale factor $\gd$ by $a=-2\gd$):
$$\eqalign{
\{\int\tr FU_2,\int\tr GU_2\}=a \int\tr\Big(
&-{1\over 3}[F,U_2]\d^{-1}[G,U_2] -[F,G]W_3\cr
&+{1\over 2}(F'G+GF'-FG'-G'F)U_2-2GF'''\Big)\ ,
\cr}
\eqn\uiii$$
$$\eqalign{
\{\int\tr FU_2,\int\tr GW_3\}=a \int\tr \Big(
&-{1\over 3}[F,U_2]\d^{-1}[G,W_3] -{1\over 6}[F,G]U_2^2\cr
&+(-{1\over 4}[F',G']+{1\over 2}[F'',G]+{1\over 12}[F,G'']) U_2\cr
&+(F'G+GF'-{1\over 2}FG'-{1\over 2}G'F)W_3 \Big)\ ,
\cr}
\eqn\uiv$$
$$\eqalign{
\{\int\tr F&W_3,\int\tr GW_3\}
=a \int\tr \Big(
-{1\over 3}[F,W_3]\d^{-1}[G,W_3]\cr
&-{1\over 6}[F,G](W_3U_2+U_2W_3)
+{2\over 3}(FU_2GW_3-GU_2FW_3)\cr
&+{5\over 12}(F'U_2GU_2-G'U_2FU_2)+{1\over 12}(FG'-GF')U_2^2
+{1\over 12}[F,G]U_2'U_2\cr
&+{7\over 12}[F',G']W_3-{1\over 6}[F,G]''W_3
+{1\over 12}(FG'''+G'''F-F'''G-GF''')U_2\cr
&+{1\over 8}(F''G'+G'F''-F'G''-G''F')U_2+{1\over 6}GF^{(5)} \Big)\ .
\cr}
\eqn\uv$$

One remarks that in the scalar case ($n=1$) this reduces to the
Poisson bracket version of Zamolodchikov's $W_3$-algebra, as it
should. In the matrix case however, even if $F=f\id,\ G=g\id$ (with
scalar $f,g$) this is a different algebra, i.e.
$\{\int\tr W_3(\s),\int\tr W_3(\sp)\}$ does not reduce to the
$W_3$-algebra, since the r.h.s. contains the non-linear terms and
$\tr U_2^2\ne (\tr U_2)^2$. In other words, the scalar ($n=1$)
$W_m$-algebras are not subalgebras of the matrix $V_{n,m}$-algebras.
The only exception is $m=2$, since one always has a Virasoro subalgebra.

In the final section 6, I first discuss various restrictions, like
imposing hermiticity conditions on the $W_k$, or for $n=2, m=2$ how
the restriction $\tr \s_3 U_2=0$ is imposed. I also mention
restrictions like setting all $U_k$ with odd $k$ to zero, etc. Some
other problems like free field representations and quantization
are briefly addressed, before I conclude.

Appendix A gives some results on pseudo-differential operators, while
in appendix B, I evaluate certain sums of products of binomial
coefficients needed in section 2.

{\chapter{The \GD brackets and the $V_{n,m}$-algebras}}

\section{Preliminaries}

In this section, I will compute the (first and) second \GD bracket of
two functionals $f$ and $g$ of the $n\times n$ matrix coefficients
$U_k(\s)$ of the linear $m^{\rm th}$-order differential operator\foot{
Throughout this paper, $m$ will denote the order of $L$ which is a
positive integer.}
$$L=-\d^m+\sum_{k=1}^mU_k\d^{m-k}\equiv \sum_{k=0}^mU_k\d^{m-k}\ .
\eqn\di$$
where $\d={d\over d\s}$.
To make subsequent formula more compact, I formally introduced
$U_0=-\id$. The fuctionals $f$ and $g$ one considers are of the form
$f=\int\tr P(U_k)$, where $P$ is some polynomial in the $U_k,\,
k=1,\ldots m$, and their derivatives (i.e. a differential polynomial
in the $U_k$). $P$ may also contain other constant or non-constant
numerical matrices so that these functionals are fairly general. (Under
suitable boundary conditions, {\it any} functional of the $U_k$ and their
derivatives can be approximated to arbitrary ``accuracy" by an $f$ of
the type considered.) The integral can either be defined in a formal
sense as assigning to any function an equivalence class by considering
functions only up to total derivatives (see e.g. section 1 of ref.
\GELF), or in the standard way if one restricts the integrand, i.e. the
$U_k$, to the class of e.g. periodic functions or sufficiently fast
decreasing functions on ${\bf R}$, etc. All that matters is that the
integral of a total derivative vanishes and that one can freely
integrate by parts.

To define the \GD brackets, it is standard to use pseudo-differential
operators
[\ADLER,\LEBMAN] involving integer powers of $\d^{-1}$. Again, $\d^{-1}$
can be defined in a formal sense by $\d \d^{-1}=\d^{-1}\d=1$, but one can
also give a concrete definition on appropriate classes of functions.
For example for $C^\infty$-functions $h$ on ${\bf R}$ decreasing
exponentially fast as $\s\to\pm\infty$ one can simply define
$(\d^{-1}h)(\s)=\int_{-\infty}^\s \rmd \sp
h(\sp)=\int_{-\infty}^\infty \rmd \sp \theta(\s-\sp)h(\sp)$.
Alternatively any of the choices
$(\d^{-1}h)(\s)=\int_{-\infty}^\infty \rmd \sp
(\l \theta(\s-\sp)-(1-\l)\theta(\sp-\s)) h(\sp)$ is equally valid,
the most symmetric choice being $\l={1\over 2}$:
$(\d^{-1}h)(\s)=\int_{-\infty}^\infty \rmd \sp
{1\over 2}\es h(\sp)$. For periodic functions on the circle,
$\d^{-1}$ is well defined on functions with no constant Fourier mode.
Then $(\d^{-1}h)(\s)=\int_{-\pi}^\pi \rmd \sp
{1\over 2}\es h(\sp)$ with $\es={1\over i\pi}\sum_{k\ne 0}{1\over k}
e^{ik(\s-\sp)}$. From the defintion of $\d^{-1}$ as inverse of $\d$
one deduces the basic formula
$$\d^{-r}h=\sum_{s=0}^\infty (-)^s\ \bin r+s-1 s h^{(s)} \d^{-r-s}\ .
\eqn\dii$$

For a pseudo-differential operator $A=\sum_{i=-\infty}^l a_i\d^i$ one
denotes
$$A_+=\sum_{i=0}^l a_i\d^i\quad ,
\quad A_-= \sum_{i=-\infty}^{-1} a_i\d^i\quad ,
\quad \res A=a_{-1}
\eqn\diii$$
so that $A=A_++A_-$. A well-known important property [\ADLER] is that for
any two pseudo-differential operators $A$ and $B$ the residue of the
commutator is a total derivative and hence $\int\res [A,B]=0$. This
is true if the $a_i, b_j$ commute with each other. In the case of
present interest the $a_i, b_j$ are matrix-valued functions and one
has instead
$$\int\tr\res [A,B]=0\ .
\eqn\div$$
For completeness, this and other properties of pseudo-differential
operators are proven in appendix A.

\section{The \GD brackets for general $U_1, \ldots U_m$}

In analogy with the scalar case (i.e. $n=1$)
[\ADLER,\LEBMAN,\GELDB,\DIK], I define the first and second \GD brackets
associated with the $n\times n$-matrix $m^{\rm th}$-order differential
operator $L$ as\foot{ To avoid confusion, let me insist that $[L,X_f]_+$
means the differential operator part of the commutator, and has nothing
to do with an anticommutator.}
$$\eqalign{
\{f,g\}_{(1)}&=a\int\tr\ \res \left( [L,X_f]_+ X_g\right)\cr
\{f,g\}_{(2)}&=a\int\tr\ \res \left( L(X_f L)_+ X_g -(L X_f)_+ L X_g
\right)  \cr }
\eqn\dv$$
where $a$ is an arbitrary scale factor and $X_f, X_g$ are the
pseudo-differential operators
$$\eqalign{
X_f=\sum_{l=1}^m \d^{-l}X_l\quad &, \quad X_g=\sum_{l=1}^m
\d^{-l}Y_l\cr
X_l= \dd f {U_{m+1-l}} \quad &, \quad Y_l= \dd g {U_{m+1-l}} \ .\cr}
\eqn\dvi$$
The functional derivative of $f=\is\tr P(U)$ is defined as usual by
$$\left( \dd f {U_k}(\s) \right)_{ij} =
\sum_{r=0}^\infty \left( -{d\over d\s}\right)^r
\left( {\d \tr P(u)(\s)\over \d (U_k^{(r)})_{ji} }\right)
\eqn\dvii$$
where $(U_k^{(r)})_{ji}$ denotes the $(j,i)$ matrix element of $r^{\rm
th}$ derivative of $U_k$. It is easily seen, that for $n=1$, equations
\dv-\dvii\ reduce to the standard
 definitions of the \GD brackets [\ADLER,\LEBMAN,\GELDB,\DIK].
 For $m=2,\, n=2$ and
with the extra restrictions $U_1=0,\ \tr\s_3U_2=0$, the second equation
\dv\  was shown in ref. \MCK\ to reproduce the original $V$-algebra \ui\
(with $a=-2\gd$).

To start with, I compute the first \GD bracket. Inserting the
definitions of $L, X_f$ and $X_g$ into the first equation \dv, and
using formula \dii\ and the cyclic commutativity \div\ under
$\int\tr {\rm res}$, it is rather staightforward to obtain
$$\eqalign{
(X_fL)_+&=\sum_{l=1}^m \sum_{k=l}^m \sum_{s=0}^{k-l} (-)^s
\bin l+s-1 s (X_lU_{m-k})^{(s)}\d^{k-l-s}\cr
(LX_f)_+&=\sum_{l=1}^m \sum_{k=l}^m \sum_{s=0}^{k-l}
\bin k-l s U_{m-k} X_l^{(s)}\d^{k-l-s}\cr}
\eqn\dviii$$
and after changing the summation indices and integrating by parts
$$
\{f,g\}_{(1)}=a\int\tr\sum_{p=0}^{m-1} \sum_{q=0}^{m-1-p}\,
\sum_{s=0}^{m-1-p-q} \bin p+s s
\left( Y_{q+1}^{(s)}X_{p+1}- X_{q+1}^{(s)}Y_{p+1} \right)
U_{m-1-p-q-s}
\eqn\dix$$
where $X_l$ and $Y_k$ are given by \dvi. After renaming $X_{s+1}\to
i^sF_s,\ Y_{r+1}\to i^rG_r,\ U_{m-k}\to (-i)^k U_k$, eq. \dix\
coincides, up to an irrelevant overall factor, with the Poisson
bracket defined by eq. 8 of ref. \GELF. It is obvious from the r.h.s.
of \dix\ that it is antisymmetric under $f\leftrightarrow g$. It is
however non-trivial to prove the Jacobi identity. This was done in
ref. \GELF.

Next, I will consider the second \GD bracket as defined by  the
second equation \dv. Contrary to the first bracket \dix\ which is
linear in the $U_k$, the second bracket, in general, will turn out to
be non-linear in the $U_k$ and will show a richer structure. First,
one has the following

\Lemma 2.1   Let
$$V_{X_f}(L)=L(X_fL)_+-(LX_f)_+L
\eqn\dx$$
which is a differential operator of order $2m-1$ at most. Define its
coefficients $\v_j$ by
$$V_{X_f}(L)=\sum_{j=0}^{2m-1}\v_j\d^j \ .
\eqn\dxi$$
Then
$$
\{f,g\}_{(2)}=a\int\tr \sum_{k=0}^{m-1} \v_k Y_{k+1} \ .
\eqn\dxii$$

\Proof The Lemma follows obviously from the definitions \dv\ of the
second \GD bracket and \dvi\ of $X_j$ together with Lemma A.2.

\noindent
Note that only the $\v_k$ with $k=0,\ldots m-1$ appear. I will show
below (Lemma 2.5) that all $\v_k$ with $k\ge m$ actually vanish, so
that $V_{X_f}$ is a mapping from the space of $m^{\rm th}$-order
differential operators onto itself. Next one has the

\Lemma 2.2
$$L(X_fL)_+=\sum_{j=0}^{2m-1} \sum_{l=1}^m \sum_{p=0}^{2m-j-l}
\sum_{q=\max(0,p+j+l-m)}^{\min(m,p+j+l)} S_{q-p,l}^{q,j}\,
U_{m-q}(X_lU_{m-p-j-l+q})^{(p)} \d^j
\eqn\dxiii$$
where for $l\ge 1$
$$
S_{r,l}^{q,j}=\sum_{s=\max(0,r)}^{\min(q,j)} (-)^{s-r}
\bin s-r+l-1 {l-1} \bin q s
\eqn\dxiv$$
with
$S_{r,l}^{q,j}=0$ if $\max(0,r)>\min(q,j)$.

\Proof Starting with \dviii\ for $(X_fL)_+$ and the (second)
definition \di\ of $L$ one has
$$L(X_fL)_+=\sum_{q=0}^{m} \sum_{l=1}^m \sum_{k=l}^{m}
\sum_{s=0}^{k-l} (-)^s \bin l+s-1 s \sum_{r=0}^q \bin q r
U_{m-q}(X_lU_{m-k})^{(s+q-r)}\d^{k-l-s+r}\ .
\eqn\dxv$$
Reshuffling the summation indices ($\tilde s=k-l-s$ and then
$j=r+\tilde s$) gives for the r.h.s of \dxv
$$\sum_{q=0}^{m} \sum_{k=0}^m \sum_{l=1}^{k}
\sum_{j=0}^{k+q-l} S_{j-k+l,l}^{q,j}
U_{m-q}(X_lU_{m-k})^{(k-l+q-j)}\d^{j}\ .
\eqn\dxvi$$
Since $S_{j-k+l,l}^{q,j}=0$ for $k<l$ one can extend the sum over $l$
from $1$ to $k$ while $k=0$ does not contribute. Also, the sum over
$j$ can be rewritten as $\sum_{j=0}^{2m-1}$ at the expense of
restricting the sum over $q$. Introducing $p=k+q-j-l$ one finally
arrives at \dxiii.

\Lemma 2.3
$$(LX_f)_+L=\sum_{j=0}^{2m-1} \sum_{l=1}^m \sum_{p=0}^{2m-j-l}
\sum_{q=\max(0,p+j+l-m)}^{\min(m,p+j+l)} \bin q-l p
U_{m-q}(X_lU_{m-p-j-l+q})^{(p)} \d^j \ .
\eqn\dxvii$$

\Proof Once again one starts with eq. \dviii, this time for
$(LX_f)_+$. However, it is more convenient to rewrite it as
$$
(LX_f)_+=\sum_{l=1}^m \sum_{k=l}^m  U_{m-k}
\d^{k-l} X_l
\eqn\dxviii$$
so that
$$(LX_f)_+L=\sum_{l=1}^m \sum_{k=l}^m \sum_{q=0}^m
\sum_{r=0}^{k-l} \bin k-l r
U_{m-k} (X_lU_{m-q})^{(k-l-r)} \d^{q+r} \ .
\eqn\dxix$$
Throughout this paper, I define a binomial coefficient $\bin a b$ to
vanish unless $a\ge b\ge 0$. Hence, the sum over $k$ can be written
$\sum_{k=0}^m$, and one then interchanges the roles of $k$ and $q$.
Introducing then $j=k+r$ and $p=q-l-r$ one finally obtains \dxvii.

Whereas \dxvii\ just contains a simple binomial coefficient, \dxiii\
contains the $S_{r,l}^{q,j}$ which are sums over  products of two
binomial coefficients. Unfortunately, for general $r,l,q,j$, I was
not able to derive a simpler expression for $S_{r,l}^{q,j}$. However,
one has the following Lemma, proven in appendix B.

\Lemma 2.4 For $q\le j$ and $r\ge l\ge 1$ one has
$$S_{r,l}^{q,j}=\bin q-l {q-r}
\eqn\dxx$$
and for $q\le j$ and $l>r\ge 0$ one has
$$S_{r,l}^{q,j}=(-)^{q-r}\bin l-1-r {q-r}
\eqn\dxxi$$

\Lemma 2.5 The coefficients of $\d^j$ in $(LX_f)_+L$ and $L(X_fL)_+$
are equal for $j\ge m$ In other words,
$$\v_j=0 \quad {\rm for}\ j\ge m\ .
\eqn\dxxii$$

\Proof   Since for any pseudo-differential operator $A$ one has
$A_+=A-A_-$ one can rewrite \dx\ as $V_{X_f}(L)=-L(X_fL)_-+(LX_f)_-L$
from which it is obvious that $V_{X_f}(L)$ is at most of degree $m-1$
and \dxxii\ follows. Alternatively, as a consistency check, it can also
be easily proven directly by considering the terms with $j\ge
m$ in eq. \dxiii\ for $L(X_fL)_+$.
 Then $q\ge
\max(0,p+j+l-m)=p+j+l-m\ge p+l$, and $q\le \min(m,p+j+l)=m\le j$. Hence
$q\le j$ and $q-p\ge l$, so that by the previous Lemma $S_{q-p,l}^{q,j}=
\bin q-l {q-(q-p)}=\bin q-l p$. Comparing with eq. \dxvii\ for
$(LX_f)_+L$ proves \dxxii\ again.

Collecting the results of Lemmas 2.1, 2.2 and 2.3 one has the

{\bf Proposition 2.6 :} The second \GD bracket associated with the
$n\times n$-matrix $m^{\rm th}$-order differential operator $L$ as
defined by eqs. \dv\ and \dvi\ equals
$$\eqalign{
\{f,g\}_{(2)}&=a\int\tr \sum_{j=0}^{m-1} \v_j Y_{j+1} \cr
\v_j&=\sum_{l=1}^m \sum_{p=0}^{2m-j-l}
\sum_{q=\max(0,p+j+l-m)}^{\min(m,p+j+l)} \left( S_{q-p,l}^{q,j} -
\bin q-l p \right)
U_{m-q}(X_lU_{m-p-j-l+q})^{(p)}\ .\cr}
\eqn\dxxiii$$

{\bf Remark 2.7 :} It will follow from the results of the next setion
on the Miura transformation that the previous Proposition gives
a well-defined symplectic structure, i.e. that \dxxiii\ is a
well-defined Poisson bracket obeying antisymmetry and the Jacobi
identity. More precisely, I will show that the $U_k$ and thus also
the functionals $f$ and $g$ are expressible in terms of certain
$n\times n$-matrices $P_i$, $i=1,\ldots m$, and that the bracket
\dxxiii\ equals the Poisson bracket of $f$ with $g$ when computed
using the much simpler $P_i$ Poisson brcket which is manifestly
antisymmetric and obeys the Jacobi identity.

{\bf Remark 2.8 :} For $m=1$, the first \GD bracket vanishes, while
the second \GD bracket is very simple. From \dxxiii\ one has
$\{f,g\}_{(2)}=a\int \tr \v_0 Y$, where $Y\equiv Y_1=\delta g/\delta
U_1$, and also $X\equiv X_1=\delta g/\delta U_1$. Now, for $m=1$,
$\v_0$ only contains 3 terms, involving $S_{0,1}^{0,0}=1$,
$S_{1,1}^{1,0}=0$ and $S_{0,1}^{1,0}=1$, so that $\v_0=-U_1X+XU_1+X'$
and
$$\{f,g\}_{(2)}=a\int\tr \left( -[X,Y] U_1 +X'Y\right) \ .
\eqn\dxxiiia$$
This is a Poisson bracket version of the $gl(n)$ Kac-Moody algebra as
considered by Drinfeld and Sokolov [\DS]. To make this even clearer,
introduce a basis $\{T^b\}, b=1,\ldots n^2$ of the Lie algebra
$gl(n)$ of $n\times n$-matrices with $[T^a, T^b]=f^{abc}T^c$ and
define for $\s$ on the unit circle $J_r^b={1\over
(-a)}\int_{-\pi}^\pi \rmd \s\, e^{ir\s}\, \tr T^b U_1(\s)$. Then eq.
\dxxiiia\ becomes
$$
i\{J^a_r, J^b_s\}_{(2)}=i f^{abc}J^c_{r+s} +\left( {2\pi\over  -a}
\right) r \delta_{r+s,0}\, \tr T^a T^b \ .
\eqn\dxxiiib$$

\section{The \GD brackets reduced to $U_1=0$}

The problem of consistently restricting a given symplectic manifold
(phase space) to a symplectic submanifold by imposing certain constraints
$\phi_i=0$ has been much studied in the literature. The basic point is
that for a given phase space one cannot set a coordinate to a given
value (or function) without also eliminating the corresponding
momentum. More generally, to impose a constraint $\phi=0$
consistently, one has to make sure that for any functional $f$ the
bracket $\{\phi,f\}$ vanishes if the constraint $\phi=0$ is imposed
{\it after} computing the bracket. In general this results in a
modification of the original Poisson bracket.

For the first \GD bracket it is easy to see that \dix\ does not
contain any terms containing $X_m=\dd f {U_1}$ or $Y_m=\dd g {U_1}$.
Hence the first \GD bracket of $U_1$ with any functional
automatically vanishes. As a consequence, one may consistently
restrict it to the submanifold of vanishing $U_1$ simply by taking
\dix\ and setting $U_1=0$ on the r.h.s.

For the second \GD bracket the situation is slightly less trivial.
One has to impose $\{f, U_1\}\vert_{U_1=0}=0$	for all $f$. Since
$Y_m=\dd g {U_1}$, one sees from \dxxiii\ that this requires
$\v_{m-1}\vert_{U_1=0}=0$. In practice this determines $X_m$ which
otherwise would be undefined if one starts with $U_1=0$. In the
scalar case it is known [\GELDB] that $X_m$ should be determined by
$\res [L,X_f]=0$. The following two Lemmas show that this is still
true in the matrix case.

\Lemma 2.9  One has
$$\v_{m-1}=-\res [L,X_f]\ .
\eqn\dxxiv$$
\Proof  On the one hand, from the definitions of $L$ and $X_f$ one
easily obtains
$$
\res [L,X_f]=\sum_{l=1}^m [U_{m+1-l},X_l]+\sum_{l=1}^m\sum_{k=l}^m
(-)^{k-l}\bin k {l-1} (X_lU_{m-k})^{(k-l+1)}\ .
\eqn\dxxvi$$
Note the commutator term which is a new feature of the present matrix
case as opposed to the scalar case. On the other hand, from eq.
\dxxiii\ one has
$$\v_{m-1}=\sum_{l=1}^m \sum_{p=0}^{m-l+1}
\sum_{q=p+l-1}^{m} \left( S_{q-p,l}^{q,m-1} -
\bin q-l p \right)
U_{m-q}(X_lU_{q-p+1-l})^{(p)}\ .
\eqn\dxxvii$$
{}From Lemma 2.4 one knows that all terms with $p+l\le q\le m-1$
vanish. The remaining terms have either $q=p+l-1$ or $q=m$. Unless
$p=m-l+1$, one has $p+l-1\le m-1$, so that using eq. \dxxi\ and eq.
(B.7) one has
$$\eqalign{
\v_{m-1}=&\sum_{l=1}^m U_{m+1-l}X_lU_0
-\sum_{l=1}^m\sum_{p=0}^{m-l}
(-)^{p}\bin p+l-1 {l-1} U_0(X_lU_{m-p-l+1})^{(p)}\cr
+&\sum_{l=1}^m (-)^{m-l} \bin m {l-1} U_0(X_lU_0)^{(m-l+1)}\ .\cr}
\eqn\dxxviii$$
The last two terms combine into $\sum_{p=0}^{m-l+1}(\ldots)$. Upon
relabelling $k=p+l-1$, and recalling $U_0\equiv -1$, the r.h.s. of
this equation is seen to coincide, up to a minus sign,  with the r.h.s.
of eq. \dxxvi. This proves \dxxiv.

\Lemma 2.10  For $U_1=0$, $\res[L,X_f]=0$ is equivalent to
$$X_m={1\over m} \sum_{l=1}^{m-1}\left(
\d^{-1}[U_{m+1-l},X_l]
+\sum_{k=l}^m  (-)^{k-l}\bin k {l-1}
(X_lU_{m-k})^{(k-l)}\right) \ .
\eqn\dxxix$$

\Proof  In eq. \dxxvi, separate the terms with $l=m$ from those with
$l\ne m$. For $U_1=0$, the terms with $l=m$ are simply $-mX_m'$ while
those with $l\ne m$ coincide with $m$ times the derivative of the
r.h.s. of \dxxix.

\noindent
One sees that the non-local term $\d^{-1}[U_{m+1-l},X_l]$ originates
from the non-commutativity of matrices and the necessity of solving for
$X_m$ when reducing to the symplectic submanifold with $U_1=0$.

The main result of this section is the following

\noindent
{\bf Theorem 2.11 :}  The second \GD bracket for $n\times n$-matrix
$m^{\rm th}$-order differential operators $L$ with vanishing $U_1$ is
given by
$$\eqalign{
\{f,g\}_{(2)}=&a\int\tr \sum_{j=0}^{m-2} \vt_j Y_{j+1}\ ,  \cr
\vt_j=&{1\over m} \sum_{l=1}^{m-1} [U_{m-j},\d^{-1} [X_l,U_{m-l+1}]]
\cr
+&{1\over m} \sum_{l=1}^{m-1}\Big\{ \sum_{k=0}^{m-l} (-)^k
\bin k+l {l-1} (X_lU_{m-k-l})^{(k)} U_{m-j} \cr
&\phantom{{1\over m} \sum_{l=1}^{m-1}} -\sum_{k=0}^{m-j-1}
\bin k+j+1 j  U_{m-k-j-1} (U_{m-l+1}X_l)^{(k)} \Big\} \cr
+&\sum_{l=1}^{m-1}\,  \sum_{p=0}^{2m-j-l}
\sum_{q=\max(0,p+j+l-m)}^{\min(m,p+j+l)} C_{q-p,l}^{q,j}
U_{m-q}(X_lU_{m-p-j-l+q})^{(p)}\ , \cr
C_{q-p,l}^{q,j}=&S_{q-p,l}^{q,j} -
\bin q-l p -{1\over m}(-)^{q-p+j}\bin q j  \bin p-q+j+l {l-1}\cr }
\eqn\dxxx$$
where the $S_{q-p,l}^{q,j}$ are defined by eq. \dxiv, and it is
understood that $U_0=-1$ and $U_1=0$.

\Proof  As discussed above, the bracket for $U_1=0$ is obtained from
the unrestricted one, eq. \dxxiii, by determining (the otherwise
undetermined) $X_m$ from $\v_{m-1}\vert_{U_1=0}=0$. This ensures that
all brackets $\{f,U_1\}$ vanish when $U_1$ is set to zero after
computing the bracket. By Lemma 2.9, $\v_{m-1}\vert_{U_1=0}=0$
implies $\res [L,X_f]\vert_{U_1=0}=0$ which by Lemma 2.10 determines
$X_m$ as given by \dxxix. All one has to do is to insert \dxxix\ into
\dxxiii. The terms $\sim X_l$ with $l<m$ are not affected and give
those terms in \dxxx\ that do not have a factor ${1\over m}$ in
front. On the other hand, the terms $\sim X_m$ in $\v_j$ are ($j<m$)
$$
 \sum_{p=0}^{m-j}
\sum_{q=p+j}^{m} \left( S_{q-p,m}^{q,j} -
\bin q-m p \right)
U_{m-q}(X_mU_{q-p-j})^{(p)}\ .
\eqn\dxxxi$$
Now, $S_{q-p,m}^{q,j}$ is non-zero only if $q-p\le \min(q,j)$. Since
$q\ge p+j$ this is only possible if $q=p+j$ in which case
$S_{j,m}^{p+j,j}=\bin p+j j$. Hence \dxxxi\ equals
$$[X_m,U_{m-j}]-\sum_{p=1}^{m-j} \bin p+j j U_{m-p-j}X_m^{(p)} \ .
\eqn\dxxxii$$
Inserting eq. \dxxix\ for $X_m$ it is straightforward to see that,
upon rearranging the summation indices, one obtains exactly all the
terms in eq. \dxxx\ for $\vt_j$ that contain a factor ${1\over m}$. One
can check again that $\vt_{m-1}$ vanishes.

\noindent
{\bf Remark 2.12 :} If one takes $m=2$, $L=-\d^2+U$, so that
$U_2\equiv U$ and $X_1\equiv X$, only $\vt_0$ is non-vanishing:
$$\vt_0=-{1\over 2} [U,\d^{-1}[U,X]]+{1\over 2}(XU+UX)'
+{1\over 2}(X'U+UX')-{1\over 2}X'''
\eqn\dxxxiii$$
and with $X=\dd f U$ and $Y=\dd g U$ one obtains (using $\int x
\d^{-1} y=-\int (\d^{-1}x) y$)
$$\{f,g\}_{(2)}=a\int\tr \left( -{1\over 2} [U,X]\d^{-1}[U,Y]
+{1\over 2}(X'Y+YX'-XY'-Y'X)U-{1\over 2}YX'''\right)
\eqn\dxxxiv$$
which obviously is a generalization of the original $V$-algebra \uii\
to arbitrary $n\times n$-matrices $U\equiv U_2$.
To appreciate the structure of the non-local terms, I explicitly
write this algebra in the simplest case for $n=2$ (but {\it without}
the restriction\foot{
see section 6.2 for a discussion of the reduction to $\tr\s_3U=0$}
$\tr\s_3U=0$ which is present for \ui). Let
$$U=\pmatrix{ T+V_3&-\rd\vp\cr -\rd\vm& T-V_3\cr} \ .
\eqn\dxxxv$$
Then one obtains from \dxxxiv\ (with $a=-2\gd$) the algebra
$$\eqalign{
\gmd \{T(\s)\, ,\, T(\s')\}  &=
(\d_\s-\d_{\s'})\left[ T(\s') \ds\right]-{1\over 2} \dsppp \cr
\gmd \{T(\s)\, ,\, V^\pm(\s')\}  &=
(\d_\s-\d_{\s'})\left[ V^\pm(\s') \ds\right]\cr
\gmd \{T(\s)\, ,\, V_3(\s')\}  &=
(\d_\s-\d_{\s'})\left[ V_3(\s') \ds\right]\cr
\gmd \{V^\pm(\s)\, ,\, V^\pm(\s')\}&=\es
V^\pm(\s)V^\pm(\s')\cr
\gmd \{V^\pm(\s)\, ,\, V^\mp(\s')\}&=-\es
(V^\pm(\s)V^\mp(\s')+V_3(\s)V_3(\s'))\cr
&\phantom{=}+(\d_\s-\d_{\s'})\left[ T(\s') \ds\right] -{1\over 2}
\dsppp \cr
\gmd \{V_3(\s)\, ,\, V^\pm(\s')\}&=\es
V^\pm(\s)V_3(\s')\cr
\gmd \{V_3(\s)\, ,\, V_3(\s')\}&=\es
(V^+(\s)V^-(\s')+ V^-(\s)V^+(\s'))\cr
&\phantom{=}+(\d_\s-\d_{\s'})\left[ T(\s') \ds\right] -{1\over 2}
\dsppp \ .\cr
}
\eqn\dxxxvi$$
Once again, one sees that $T$ generates the conformal algebra with a
central charge, while $V^+$, $V^-$ and $V_3$ are conformally primary
fields of weight (spin) two. It is easy to check on the example \dxxxiv,
that antisymmetry and the Jacobi identity are satisfied. For general $m$
this follows from the Miura transformation to which I now turn.

\chapter{{The Miura transformation}}

In this section, I will give the matrix Miura transformation. By this
transformation, all $U_k(\s)$ can be expressed as differential
polynomials in certain $ n\times n$-matrices $P_j(\s), j=1,\ldots m$,
and hence every functional $f$ of the $U_k$ will also be a functional
$\tilde f$ of the $P_j$. I will define a very simple Poisson bracket for
functionals of the $P_j$. Using this Poisson bracket one can compute
$\{\tilde f(P),\tilde g(P)\}\equiv \{f(U(P)),g(U(P))\}$.
I will show that this Poisson bracket coincides with the second \GD
bracket $\{f(U),g(U)\}_{(2)}$ defined by eq. \dxxiii\ in the previous
section. As a corollary, this proves the antisymmetry and Jacobi
identity for the latter. The Poisson bracket of the $P_j$ can be
reduced to the submanifold where $\sum_{j=1}^mP_j=0$. This implies
$U_1=0$. Then I will show that this reduced Poisson bracket for the
$P_j$ gives the second \GD bracket for the $U_k$ reduced to $U_1=0$,
i.e. eq. \dxxx.

\section{The case of general $U_1, \ldots U_m$}

{\bf Definition and Lemma 3.1 :}  Introduce the $n\times
n$-matrix-valued functions $P_j(\s),\, j=1, \ldots m$. Then for
functionals $f,g$ (integrals of traces of differential polynomials)
of the $P_j$ the following Poisson bracket is well-defined %
$$\{f,g\}=a\sum_{i=1}^m \int\tr\left( \left( \dd f {P_i} \right)'
\dd g {P_i} - \left[ \dd f {P_i} , \dd g {P_i} \right] P_i \right)
\eqn\ti$$
or equivalently for $n\times n$-matrix-valued (numerical) functions
$F$ and $G$
$$\{\int\tr FP_i,\int\tr GP_j\}=a\,  \delta_{ij} \int\tr \left( F'G
-[F,G]P_i \right) \ .
\eqn\tii$$

\Proof The definition \ti\ is obviously bilinear in $f$ and $g$ and
reduces to \tii\ for the special functionals chosen. When \tii\ is
chosen as starting point, bilinearity is implicitly understood, so
that \ti\ follows. Antisymmetry is obvious for both \ti\ and \tii. It
is easy to see that the Jacobi identity for the bracket \tii\ amounts
to the Jacobi identity for the matrix commutator $[F,G]$ which, of
course, is satisfied.

\noindent
Note that due to the $\delta_{ij}$ in \tii\ one has $m$ decoupled
Poisson brackets. In the scalar case ($n=1$), \tii\ simply gives
$\{P_i(\s),P_j(\sp)\}=(-a)\delta_{ij}\dsp$. These are $m$ free fields
or $m$ $U(1)$ current algebras. In the matrix case, comparing \ti\ or
\tii\ with \dxxiiia\ one sees that each $P_j$ actually gives a $gl(n)$
current algebra. So one has no lnger free  fields but $m$ completely
decoupled current algebras. This is still much simpler than the
bracket \dxxiii. To connect both brackets one starts with the
following obvious

\Lemma 3.2  Let $P_j, j=1, \ldots m$ be as in Lemma 3.1. Then
$$L=-(\d-P_1)(\d-P_2)\ldots (\d-P_m)
\eqn\tiii$$
is a $m^{\rm th}$-order $n\times n$-matrix linear differential
operator and can be written $L=\sum_{k=0}^mU_{m-k}\d^k$ with $U_0=-1$
as before. This identification gives all $U_k, k=1,\ldots m$ as
$k^{\rm th}$-order differential polynomials in the $P_j$, i.e. it
provides an embedding of the algebra of differential polynomials in
the $U_k$ into the algebra of differential polynomials in the $P_j$.
One has in particular
$$U_1=\sum_{j=1}^m P_j\quad , \quad
U_2=-\sum_{i<j}^m P_iP_j+\sum_{j=2}^m (j-1)P_j'\ .
\eqn\tiv$$

\noindent
I will call the embedding given by \tiii\ a (matrix) Miura
transformation. The most important property of this Miura
transformation is given by the following matrix-generalization of a
well-known theorem [\KW, \DIK].

{\bf Theorem 3.3 :}  Let $f(U)$ and $g(U)$ be functional of the $U_k,\
k=1,\ldots m$. By Lemma 3.2 they are also functionals of the $P_j,\
j=1,\ldots m$: $f(U)=\ft(P)$, $g(U)=\gt(P)$ where $\ft(P)=f(U(P))$
etc. One then has
$$\{\ft(P),\gt(P)\}=\{f(U),g(U)\}_{(2)}
\eqn\tv$$
where the bracket on the l.h.s. is the Poisson bracket \ti\ and the
bracket on the r.h.s. is the second \GD bracket \dxxiii.

\Proof  In the scalar case, the simplest proof is probably the one
given by Dikii [\DIK]. Here, I repeat this proof, adapting it to the
matrix case were necessary. The l.h.s. of \tv\ is given by the r.h.s.
of \ti\ which can be rewritten as
$$\{\ft(P),\gt(P)\}=a\sum_i\int\tr \dd {\gt } {P_i} \left[ \d-P_i,
\dd {\ft } {P_i} \right]
\eqn\tvi$$
whereas the r.h.s. of \tv\ is given by \dxxiii\ which I recall is
the explicit form of $a\int\tr\res (L(X_fL)_+X_g -(LX_f)_+LX_g)$.
Since $X_f$ and $X_g$ contain $\dd f {U_k}$ and $\dd g {U_k}$ one needs
to obtain $\dd {\ft } {P_i}$ in terms of $\dd f {U_k}$, and similary for
$g$. This is done as follows. One has, using $\delta L=\sum_{k=1}^m
\delta U_k \d^{m-k}$, and $X_f=\sum_{l=1}^m \d^{-l} \dd f {U_{m+1-l}}$,
as well as Lemma A.1
$$\int\tr\res X_f \delta L = \int\tr\sum_{k=1}^m \dd f {U_k} \delta U_k
= \delta f(U)\ .
\eqn\tvii$$
On the other hand, denoting $\d_i=\d-P_i$, so that $L=-\d_1\ldots\d_m$,
one has  $\delta L=\sum_{i=1} \d_1\ldots \d_{i-1} \delta P_i
\d_{i+1}\ldots \d_m$ so that using Lemmas A.1 and A.3
$$\int\tr\res X_f \delta L = \int\tr \sum_{i=1}^m \delta P_i\, \res
(\d_{i+1}\ldots \d_mX_f\d_1\ldots \d_{i-1})\ .
\eqn\tviii$$
Writing $\delta f(U)=\delta \ft(P)=\int\tr\sum_{i=1}^m\delta P_i
\dd {\ft } {P_i}$ one obtains from \tvii\ and \tviii\ the desired
expression of $\dd {\ft } {P_i}$ in terms of $\dd f {U_k}$:
$$\dd {\ft } {P_i} = \res (\d_{i+1}\ldots \d_mX_f\d_1\ldots \d_{i-1})\ .
\eqn\tix$$
Then eq. \tvi\ becomes
$$\eqalign{
&\{\ft(P),\gt(P)\}\cr
&=a\sum_i\int\tr
\res (\d_{i+1}\ldots \d_mX_g\d_1\ldots \d_{i-1})
\left[ \d_i,
\res (\d_{i+1}\ldots \d_mX_f\d_1\ldots \d_{i-1})
 \right] \ . \cr}
\eqn\tx$$
At this point one has an important difference with the scalar case.
In the scalar case, eq. \tvi\ reads $a\sum_i\int \dd {\gt } {P_i} \left(
\dd {\ft } {P_i} \right)'$ so that only $\d$ not $\d_i\equiv \d-P_i$
appears in the commutator with $\res(\ldots)$ in \tx. But in the scalar
case, $P_i$ commutes with $\res(\ldots)$, so that one could replace $\d$
by $\d_i$ and also obtain \tx. In the present matrix case, however, one
cannot simply replace $\d$ by $\d_i$, and one needs $\d_i$ from the
beginning in \tvi. The rest of the proof goes as in the scalar case
[\DIK] and I only give it here for completeness. Using first Lemmas
A.3 and A.1, and then Lemma A.4, the r.h.s. of \tx\ becomes
$$\eqalign{
a\sum_i\int\tr
\res &\big(\d_{i+1}\ldots \d_mX_g\d_1\ldots \d_{i}\,
\res (\d_{i+1}\ldots \d_mX_f\d_1\ldots \d_{i-1})  \cr
&-\d_{i}\ldots \d_mX_g\d_1\ldots \d_{i-1}\,
\res (\d_{i+1}\ldots \d_mX_f\d_1\ldots \d_{i-1})\big)  \cr
=a\sum_i\int\tr
\res &\big(\d_{i+1}\ldots \d_mX_g\d_1\ldots \d_{i}
\big( (\d_{i+1}\ldots \d_mX_f\d_1\ldots \d_{i-1})_- \d_i\big)_+  \cr
&-\d_{i}\ldots \d_mX_g\d_1\ldots \d_{i-1}
\big( \d_i (\d_{i+1}\ldots \d_mX_f\d_1\ldots \d_{i-1})_-\big)_+ \big)
\ . \cr
}
\eqn\txi$$
If one would replace the $-$ subscripts by $+$ subscripts, this
expression would vanish (since then the external $+$ subscripts could
be dropped and both terms cancel). Since
$(\ldots)_- =(\ldots)-(\ldots)_+$, one can thus simply drop the $-$
subscripts to obtain
$$\eqalign{
&\{\ft(P),\gt(P)\}=a\sum_{i=1}^m (S_{i+1}-S_i)\ , \cr
&S_i=\int\tr
\res \d_{i}\ldots \d_mX_g\d_1\ldots \d_{i-1}
( \d_i \ldots \d_mX_f\d_1\ldots \d_{i-1})_+ \ .\cr }
\eqn\txii$$
Performing the sum, all $S_i$ cancel except for
$$\eqalign{
S_{m+1}=\int\tr \res X_g\d_1\ldots \d_{m}
(X_f\d_1\ldots \d_{m})_+ =\int\tr\res X_gL(X_fL)_+ \ ,\cr
S_1=\int\tr \res \d_1\ldots \d_{m}X_g
(\d_1\ldots \d_{m}X_f)_+ =\int\tr\res LX_g(LX_f)_+ \ ,\cr
}
\eqn\txiii$$
so that
$$\{\ft(P),\gt(P)\}=a(S_{m+1}-S_1) =a \int\tr\res (L(X_fL)_+X_g
-(LX_f)_+LX_g)
\eqn\txiv$$
which completes the proof.

\noindent
The previous proposition states that one can either compute
$\{U_k,U_l\}$ using the complicated formula \dxxiii\ or using the
simple Poisson bracket \ti\ for more or less complicated functionals
$U_k(P)$ and $U_l(P)$. In particular Lemma 3.1 implies the

\noindent
{\bf Corollary 3.4 :} The second \GD bracket \dxxiii\ obeys
antisymmetry and the Jacobi identity. Bilinearity in $f$ and $g$
being evident, it is a well defined Poisson bracket.

\section{The case $U_1=0$}

As seen from \tiv, $U_1=0$ corresponds to $\sum_{i=1}^m P_i=0$. In
order to describe the reduction to $\sum_i P_i=0$ it is convenient to
go from the $P_i, i=1,\ldots m$ to a new ``basis": $Q=\sum_{i=1}^m
P_i$ and $\P_a, a=1,\ldots m-1$ where all $\P_a$ lie in the
hyperplane $Q=0$. Of course, $Q$ and each $\P_a$ are still $n\times
n$-matrices. More precisely:

{\bf Definition and Lemma 3.5 :}  Consider a $(m-1)$-dimensional
vector space, and choose an overcomplete basis of $m$ vectors $h_j,\
j=1,\ldots m$. Denote the components of each $h_j$ by $h_j^a,\  a=1,
\ldots m-1$. Choose the $h_j$ be such that
$$\eqalign{
\sum_{j=1}^m h_j &=0\ ,\cr
h_i\cdot h_j &= \delta_{ij}-{1\over m}\ ,\cr
\sum_{i=1}^m h_i^a h_i^b &=\delta_{ab} \cr}
\eqn\txv$$
and define the completely symmetric rank-3 tensor $D_{abc}$ by
$$D_{abc}=\sum_{i=1}^m h_i^a h_i^b h_i^c \ .
\eqn\txvi$$
Define $Q$ and $\P_a,\  a=1,\ldots m-1$ to be the following linear
combinations of the $P_j$'s
$$\P_a=\sum_{j=1}^m h_j^a P_j \quad , \quad Q=\sum_{j=1}^m P_j\ .
\eqn\txvii$$
If one considers the $P_j$ as an orthogonal basis in a $m$-dimensional
vector-space, then the $\P_a$ are an orthogonal basis in a
$(m-1)$-dimensional hyperplane orthogonal to the line spanned by $Q$.
Equation \txvii\ is inverted by
$$P_i={1\over m}Q+\sum_{a=1}^{m-1} h_i^a \P_a \ .
\eqn\txviii$$

\Proof  First, note that the vectors $h_j$ with the desired properties
\txv\ exist, since one can choose them to be the weight vectors of the
vector representation of $SU(m)\sim A_{m-1}$. Next, considering the
$P_i$ as orthogonal means that one formally introduces some inner
product $(P_i,P_j)=\delta_{ij}$. Then obviously from the definitions
\txvii\ and the properties \txv\ of the $h_j$ one has $(\P_a, \P_b)
=\delta_{ab}$, as well as $(Q,\P_a)=0$. Finally, eq. \txviii\ also is
an immediate consequence of \txvii\ and \txv.

\noindent
Thus it is convenient to use $Q$ and the $\P_a$ to discuss the
reduction to $Q=0$.  Note that the occurrence of the weights of the
vector representation of $A_{m-1}$ is not surprising since in the
scalar case, $n=1$, the reduction to $Q=0$ is well-known to be
related to $A_{m-1}$ via the Drinfeld-Sokolov reduction [\DS].

By Lemma 3.5 any functional of the $P_i$ can be written as a
functional of the $\P_a$ and $Q$. One has

\Lemma 3.6   If for functionals $f(\P,Q),\,  g(\P,Q)$ one denotes
$$V_a= \dd f {\P_a} \quad , \quad V_0= \dd f Q \quad , \quad
W_a= \dd g {\P_a} \quad , \quad W_0= \dd g Q
\eqn\txix$$
then the Poisson bracket \ti\ of $f$ with $g$ is\foot{
One should not confuse the scale factor $a$ in front of the integral
sign with the vector indices $b,c,d$.}
$$\eqalign{
\{f,g\}=a\int\tr &\Big( mW_0V_0' + W_bV_b' - [V_0,W_0]Q -{1\over m}
[V_b,W_b] Q\cr
&-[V_0,W_b]\P_b-[V_b,W_0]\P_b -D_{bcd}[V_b,W_c]\P_d \Big) \cr}
\eqn\txx$$
where here and in the following repeated vector indices ($b,c,d$) are
to be summed over.

\Proof   From \txvii\ one has $\dd f {P_i} =\dd f Q +h_i^a \dd f {\P_a}
=V_0+h_i^aV_a$ and similarly $\dd g {P_i} =W_0+h_i^b W_b$. Inserting
these relations into \ti\ and using relations \txv, \txvi\ and
\txviii\ yields \txx.

{\bf Proposition 3.7} The Poisson bracket \ti\ can be reduced to the
symplectic submanifold with $Q\equiv \sum_{j=1}^m P_j =0$. The reduced
Poisson bracket is
$$\{f,g\}=a\int\tr \Big(  W_bV_b'
-[V_b,W_c]D_{bcd}\P_d -{1\over m} [V_b,\P_b]\d^{-1}[W_c,\P_c]\Big)
\eqn\txxi$$
where $V_b, W_b$ are defined in eq. \txix, or equivalently
$$\{\int\tr F\P_b,\int\tr G\P_c\}=a\int\tr \Big(  GF'\delta_{bc}
-[F,G]D_{bcd}\P_d -{1\over m} [F,\P_b]\d^{-1}[G,\P_c]\Big)
\eqn\txxii$$

\Proof   The reduced bracket is obtained from the unreduced one, eq.
\txx, by determining $V_0$ and $W_0$ such that the Poisson bracket of
$Q$ with any functional $g$ vanishes on the submanifold $Q=0$. From
\txx\ one has
$$
\{\int\tr V_0 Q,g\}=a\int\tr V_0 \big( -mW_0' +[Q,W_0]
+[\P_b,W_b] \big)
\eqn\txxiii$$
The vanishing of the r.h.s. for $Q=0$ gives
$$W_0={1\over m}\d^{-1}[\P_b,W_b]
\eqn\txxiv$$
and similarly $V_0={1\over m}\d^{-1}[\P_b,V_b]$. Inserting these
relations into \txx\ and setting $Q=0$ yields \txxi. Equation \txxii\
follows obviously from \txxi.

\noindent
Of course, the result \txxii\ can also be expressed in terms of the
$P_i$ directly. Using \txviii\ for $Q=0$, i.e. $P_i=h_i^a\P_a$, one
immediately obtains from \txxii, using the relations \txv\ and \txvi\
the

\noindent
{\bf Corollary 3.8 :}  The Poisson bracket \tii\ when reduced to
$Q=\sum_{j=1}^mP_j=0$ can be equivalently written as
$$\eqalign{
\{f,g\}=a\int\tr &\Big(  GF'(\delta_{ij} -{1\over m}) -
[F,G] (\delta_{ij} -{2\over m}){1\over 2}(P_i+P_j)\cr
&-{1\over m} [F,P_i]\d^{-1}[G,P_j]\Big) \ . \cr}
\eqn\txxv$$
To prove the main result of this section one needs the following

\Lemma 3.9   Let as before $X_f=\sum_{l=1}^m \d^{-l} X_l,\
\d_i=\d-P_i$, so that $L=-\d_1\ldots\d_m$, and let
$$\pi_i=\d_{i+1}\ldots \d_mX_f\d_1\ldots \d_{i-1}
\eqn\txxvi$$
then
$$\sum_{i=1}^m [\d_i, \res \pi_i] = \res [X_f, L]\ .
\eqn\txxvii$$

\Proof   By Lemma A.5, one has $[\d_i, \res \pi_i]=\res [\d_i,\pi_i]$
so that the l.h.s. of \txxvii\ is
$$\eqalign{
&\res \sum_{i=1}^m (\d_i\pi_i-\pi_i\d_i) = \res \sum_{i=1}^m
(\d_{i}\ldots \d_mX_f\d_1\ldots \d_{i-1}
- \d_{i+1}\ldots \d_mX_f\d_1\ldots \d_{i} )\cr
&= \res (\d_{1}\ldots \d_mX_f - X_f\d_1\ldots \d_{m})
=\res [X_f,L]\ .\cr}
\eqn\txxviii$$

\noindent
Now one has the

\noindent
{\bf Theorem 3.10 :}   Let $U_1=0$ and hence $Q=\sum_{j=1}^mP_j=0$.
By the Miura transformation of Lemma 3.2 any functionals $f(U),\, g(U)$
of the $U_k, k=2,\ldots m$ only are also functionals
$\ft(\P)=f(U(\P)),\, \gt(\P)=g(U(\P))$ of the $\P_a, a=1, \ldots m$
only. The reduced second \GD bracket \dxxx\ of $f$ and $g$ equals the
reduced Poisson bracket \txxi\ of $\ft$ and $\gt$.

\Proof   Recall that for $U_1=0$ the reduced bracket \dxxx\ equals
the unreduced one $\{f,g\}_{(2)}= a\int\tr\res (L(X_fL)_+X_g
-(LX_f)_+LX_g)$ supplemented by the constraint  $\res[X_f, L]=0$ which
determines $X_m$. Recall also that for $\sum_{j=1}^mP_j=0$ the reduced
bracket \txxi\ equals the unreduced one, eq. \ti, which can be written
in the form \tvi, supplemented by the constraint that the functional
derivative $\dd g Q$ does not appear, or in other words, as seen from
\tvi, that $\sum_{i=1}^m [ \d_i, \dd {\ft} {P_i} ]=0$. By Theorem 3.3,
the unreduced brackets are equal. Hence the reduced ones will be
equal if the reducing constraints are equivalent. But since from \tix\
one has $\dd {\ft} {P_i} =\res \pi_i$, the previous Lemma shows that
$\res [X_f,L]=0$ is equivalent to
$\sum_{i=1}^m [ \d_i, \dd {\ft} {P_i} ]=0$, and the theorem follows.

\noindent
{\bf Corollary 3.11 :}  The second \GD bracket \dxxx\ obeys
antisymmetry and the Jacobi identity. Bilinearity in $f$ and $g$ being
evident, it is a well-defined Poisson bracket.

\noindent
{\bf Example 3.12 :}  Consider the example $m=2$. The $h_j$ can be
taken to be the negative of the weight vectors of $SU(2)$ which are
one-dimensional: $h_1=-{1\over \rd}, h_2={1\over \rd}$. Then there is
only one $\P$ which by \txviii\ equals $\P=\rd P_2=-\rd P_1$, and
$U\equiv U_2={1\over 2} \P^2+{1\over \rd}\P$. Since $D_{bcd}=0$ for
$m=2$, the Poisson bracket \txxii\ becomes
$$\{\int\tr F\P,\int\tr G\P\}=a\int\tr \big(  GF'
-{1\over 2} [F,\P]\d^{-1}[G,\P]\big)\ .
\eqn\txxix$$
It can be easily checked directly that this implies the second \GD
bracket \dxxxiv.

\chapter{{The conformal properties}}

In the scalar case, i.e. for $n=1$, the second \GD bracket (with
$U_1=0$) gives the $W_m$-algebras [\BAK,\MAT,\DIZ]. The interest in the
$W$-algebras stems from the fact that they are extensions of the
conformal Virasoro algebra, i.e. they contain the Virasoro algebra as a
subalgebra. Furthermore, in the scalar case, it is known that certain
combinations of the $U_k$ and their derivatives yield primary fields of
integer spins $3,4,\ldots m$. It is the purpose of this section to
establish the same results for the matrix case, $n>1$. Throughout this
section, I only consider the second \GD bracket \dxxx\ for the case
$U_1=0$. I will simply write $\{f,g\}$ instead of $\{f,g\}_{(2)}$.
Also, it is often more convenient to replace the scale factor $a$ by
$\gd$ related to $a$ by
$$a=-2\gd\ .
\eqn\qi$$
(Note that $\gd$ need not be positive.)

\section{The Virasoro subalgebra}

For the $V$-algebra \ui\ given in the introduction (corresponding to
$m=2,\, n=2$ and an additional constraint $\tr\s_3U_2=0$) one sees that
$T={1\over 2}\tr U_2$ generates the conformal algebra. I will now show
that for general $m, n$, the generator of the conformal algebra is
still given by this formula.

\Lemma 4.1   For arbitrary $m\ge 2$ one has
$$\eqalign{
\{\int\tr FU_2,\int\tr GU_2\}&
= a\int\tr \Big(
-{1\over m} [F,U_2]\d^{-1}[G,U_2]-[F,G](U_3-{m-2\over 2}U_2')\cr
&+{1\over 2}(F'G-G'F+GF'-FG')U_2 -{1\over 2} \bin m+1 3  G F''' \Big) \
. \cr}
\eqn\qii$$
Note that for $m=2$ one has to set $U_3=0$.

\Proof   From eq. \dxxx\ one knows that the l.h.s. of \qii\ equals
$a\int\tr \vt_{m-2} G$ with $f=\int\tr FU_2$, i.e.
$X_l=F\delta_{l,m-1}$. Still from \dxxx\ one has (recalling $U_1=0$,
$U_0=-1$)
$$\eqalign{
\vt_{m-2}\Big\vert_{X_l=F\delta_{l,m-1}}&=
{1\over m} [U_2, \d^{-1} [F,U_2]] +{1\over m} \bin m 2 (F'U_2+(U_2F)')
\cr
&+\sum_{p=0}^3 \, \sum_{q=\max(0,p+m-3)}^{\min(m,p+2m-3)}
C_{q-p,m-1}^{q,m-2} U_{m-q}(FU_{q-p+3-m})^{(p)} \ . \cr}
\eqn\qiii$$
At this point one needs explicit expressions for the coefficients
$C_{q-p,m-1}^{q,m-2}$, i.e. of the $S_{q-p,m-1}^{q,m-2}$. Note that the
latter are non-vanishing only if $q-p\le \min(q,m-2)$, i.e. $q\le
p+m-2$. Some of them follow from Lemma 2.4. The others have to be
obtained directly from the definition \dxiv\ which is not difficult
since the sum involves at most $(m-2)-(q-p)+1\le (m-2)-(m-3)+1=2$
terms. Since $U_1=0$ only five $C$-coefficient are needed. They are
$C_{m-3,m-1}^{m-3,m-2}=1$,
$C_{m,m-1}^{m,m-2}=-1$,
$C_{m-3,m-1}^{m-2,m-2}=C_{m-1,m-1}^{m,m-2}={m-3\over 2}$,
$C_{m-3,m-1}^{m,m-2}=-{(m+1)m(m-1)\over 12}$. Inserting this into
\qiii\ and performing some simple algebra gives \qii.

\noindent
{\bf Proposition 4.2 :}  Let $T(\s)={1\over 2} \tr U_2(\s)$. Then
$$\gmd \{T(\s_1),T(\s_2)\}=(\d_{\s_1}-\d_{\s_2})
\left( T(\s_2)\delta(\s_1-\s_2)\right)
-{n\over 4}\bin m+1 3 \delta'''(\s_1-\s_2) \ .
\eqn\qiv$$
Equivalently, if, for $\s\in S^1$, one defines for integer $r$
$$L_r=\gmd\int_{-\pi}^\pi \rmd\s\,  T(\s)e^{ir\s} +{c\over
24}\, \delta_{r,0}
\eqn\qv$$
where
$$c={6\pi\over \gd} \, n\, \bin m+1 3 = {12\pi\over (-a)} n\, \bin m+1 3
\eqn\qvi$$
then the $L_r$ form a Poisson bracket version of the Virasoro algebra
with (classical) central charge $c$ :
$$i\{L_r,L_s\}=(r-s)L_{r+s}+{c\over 12} (r^3-r) \delta_{r+s,0} \ .
\eqn\qvii$$
Also if $\{A_\mu\}_{\mu=1,\ldots n^2-1}$ is a basis for the traceless
$n\times n$-matrices, then each $S_\mu (\s)=\tr A_\mu U_2(\s),
\mu=1,\ldots n^2-1$ is a conformally primary field of conformal
dimension (spin) 2: %
$$\gmd \{T(\s_1),S_\mu(\s_2)\}=(\d_{\s_1}-\d_{\s_2})
\left( S_\mu(\s_2)\delta(\s_1-\s_2)\right)
\eqn\qviii$$
or for the modes $(S_\mu)_r=\gmd\int_{-\pi}^\pi \rmd\s\,
S_\mu(\s)e^{ir\s}$ one has equivalently
$i\{L_r,(S_\mu)_s\}=(r-s)(S_\mu)_{r+s}$. Equations \qiv\ and \qviii\
can be written in matrix notation as ($\id$ denotes the $n\times n$
unit matrix)
$$\gmd \{T(\s_1),U_2(\s_2)\}=(\d_{\s_1}-\d_{\s_2})
\left( U_2(\s_2)\delta(\s_1-\s_2)\right)
-{1\over 2}\bin m+1 3 \id\,  \delta'''(\s_1-\s_2) \ .
\eqn\qix$$

\Proof   Consider first \qii\ with $F(\s)={1\over 2} \delta(\s-\s_1)
\id$ and $G(\s)={1\over 2} \delta(\s-\s_2) \id$. Then the l.h.s. of
\qii\ is  $\{T(\s_1),T(\s_2)\}$ while on the r.h.s. all commutator
terms vanish. Recall $a=-2\gd$ and $\tr \id =n$ and eq. \qiv\ follows.
It is then standard (and straightforward) to show that \qv, \qvi\ imply
\qvii. Finally let $F(\s)={1\over 2} \delta(\s-\s_1) \id$
and $G(\s)=\delta(\s-\s_2) A_\mu$. Again, all commutator terms vanish
in \qii\ and \qviii\ follows. Eq. \qix\ then is obvious.

Note that \qvi\ is a classical central charge. If one can implement
a free-field quantization, the central charge receives additional
normal-ordering contributions expected to be $(m-1)n^2$ so that $c_{\rm
tot}=n(m-1)\left( n+{\pi\over \gd}m(m+1)\right)$. One could speculate
about series of unitary representation etc, but I will not do so here.

As a sideremark, let me note
that for the case of unrestricted $U_1$, the conformal
generator is no longer given by $\tr U_2$ but by some linear
combination of $\tr U_2$, $\tr U_1^2$ and $\tr U_1'$. The
coefficient of $\tr U_1'$ is arbitrary and appears as a parameter
in the central charge.

\section{The conformal properties of the $U_k$ for $k\ge 3$}

In the previous subsection, I have computed the conformal properties of
the matrix elements of $U_2$. The aim of this subsection is to give
those for all other $U_k$, i.e. compute $\{T(\s_1),U_k(\s_2)\}$ or
equivalently for any (test-) function $\e(\s)$ $\{\int \e T, U_k(\s_2)\}$
for all $k\ge 3$. I will find that this Poisson bracket is linear in the
$U_l$ and their derivatives and is formally identical to the result of
the scalar case. It then follows that appropriately symmetrized
combinations $W_k$ can be formed that are $n\times n$-matrices, each
matrix element of $W_k$ being a conformal primary field of dimension
(spin) $k$.

\Lemma 4.3   For a scalar function $\e$ and a $n\times n$-matrix-valued
function $F$, one has
$$\gmd\{\int\e T, \int\tr FU_k\}=
\int\e\, \tr\vt_{m-2}\Big\vert_{X_l=F\delta_{l,m+1-k}}
\eqn\qx$$
with
$$\eqalign{
&\tr\vt_{m-2}\Big\vert_{X_l=F\delta_{l,m+1-k}} =
{1\over m}\sum_{q=0}^{k-1}(-)^q\bin m-k+q+1 {m-k} \tr (FU_{k-1-q})^{(q)}
U_2 \cr
&+{m-1\over 2} \tr (U_kF)'
 + \sum_{p=0}^{k+1}\, \sum_{q=\max(0,k+1-p-m)}^{\min(m,k+1-p)}
C_{m-q-p,m+1-k}^{m-q,m-2} \tr U_q(FU_{k+1-p-q})^{(p)} \ .\cr}
\eqn\qxi$$

\Proof  The l.h.s. of \qx\ equals
$-{1\over 2\gd}\{\int\tr FU_k, \int\e\, \id T \}$ which by \dxxx\ and
\qi\ equals $\int\tr\sum_{j=0}^{m-2}\vt_j Y_{j+1}$ with $Y_{j+1}=\e\,
\id \delta_{j,m-2}$ and $X_l=F\delta_{m+1-l,k}$ which proves \qx.
Equation \qxi\ then follows upon inserting $j=m-2$ and $X_l$ into eq.
\dxxx\ for $\vt_j$ and changing the summation index in the last sum from
$q$ to $m-q$.

\Lemma 4.4  When evaluating the sums in the third term on the r.h.s.
of  \qxi\ one has to distinguish three cases:
The coefficients $C_{m-q-p,m+1-k}^{m-q,m-2}$, defined in eq. \dxxx, are
given by ($2\le k\le m$)\foot{the equations are actually also valid for
$k=1$.} \nextline
a) $q\ge 2$ and $q\le k-1-p$
$$C_{m-q-p,m+1-k}^{m-q,m-2}={1\over m} (-)^{p+1}\, \delta_{q,2}
\bin m-k+p+1 {m-k}
\eqn\qxii$$
b) $q\ge 2$ and $q= k+1-p$
$$C_{m-q-p,m+1-k}^{m-q,m-2}=(-)^p\bin 1 p + {1\over m}
(-)^{k}\, \delta_{p,k-1} \bin m k
\eqn\qxiii$$
c) $q=0$
$$C_{m-q-p,m+1-k}^{m-q,m-2}=(-)^p\  {m+1\over 2}
\bin m-k+p-1 {m-k} -(-)^p\bin m-k+p {m-k}
\eqn\qxiv$$

\Proof   First note that since $U_1=0$ no terms with $q=1$ or $q=k-p$
are present in the  sum considered in \qxi. Thus one only has to
distinguish $q=0$ or $q\ge 2$ on the one hand, and $q\le k-1-p$ or
$q=k-p+1$ on the other hand. Hence one has the cases a) and b), while in
case c) one should distinguish $p\le k-1$ and $p=k+1$. For cases a) and
b), $q\ge 2$ and $m-q\le m-2$, so that one can use Lemma 2.4 to evaluate
$S_{m-q-p,m+1-k}^{m-q,m-2}$. In case a), eq. \dxx\ applies and one gets
eq. \qxii. In case b), eq. \dxxi\ applies if also $q\le m-p$, i.e.
$k<m$, so that $S_{m-q-p,m+1-k}^{m-q,m-2}=(-)^p \bin q+p-k p =(-)^p
\bin 1 p$. If however $k=m$, so that $q=m-p+1$ (which is only possible
for $p\ge 1$), then $S_{m-q-p,m+1-k}^{m-q,m-2}=S_{-1,1}^{m-q,m-2}$
which is easily evaluated directly from the definition \dxiv\ to be
$-\delta_{q,m}=-\delta_{p,1}=(-)^p \bin 1 p$ since $p\ge 1$. Hence
\qxiii\ follows. In case c), one has $q=0$, hence $m-q>m-2$ and Lemma
2.4 does not apply directly. However, using equation (B.8) first, one can
still use Lemma 2.4, i.e. eq. \dxx\ if $p\le k-1$ and eq. \dxxi\ if
$p=k+1$ ($p=k,q=0$ gives a $U_1$ and does not contribute). For
$p\le k-1$ it is straightforward to obtain \qxiv. For $p=k+1$ one has
still to distinguish $k=m$ and $k<m$, but the result is the same in
both situations, and it differs from \qxiv\ only by a term $\bin k-1
p$ which vanishes since $p=k+1$. Hence one obtains \qxiv\ again.

\noindent
{\bf Proposition 4.5 :}
The conformal properties of all matrix elements of all $U_k, k=2,\ldots
m$ are given by
$$\eqalign{
\gmd \{\int \e T, U_k\} &= -\e U_k'-k\e' U_k +{k-1\over 2} \bin m+1 {k+1}
\e^{(k+1)}\cr
& +\sum_{l=2}^{k-1}\left[ \bin m-l {k+1-l} -{m-1\over 2} \bin m-l {k-l}
\right] \e^{(k-l+1)} U_l \cr }
\eqn\qxv$$
which is formally the same equation as in the scalar case $n=1$.

\Proof   That \qxv\ is the same equation as in the scalar case is
readily seen by comparing with eq. (2.10) of ref. \DIZ, setting $\gd=1$
and observing that the $a_l$ of ref. \DIZ\ correspond to the present
$-U_l$ (for $n=1$), and thus also $a_2$ corresponds to $-T$. Let me now
prove \qxv\ in the matrix case. Note that for $k=2$, eq. \qxv\ is
equivalent to \qix. Hence one only needs to consider $k\ge 3$. One
starts with eq. \qx\ and inserts the results of Lemma 4.4 into eq.
\qxi. Case a) can be realized if $2\le q \le k-1-p$, i.e. for $p\le k-3$.
Case b) can be realized if $p\le k-1$, while case c) can be realized if
$k+1-m\le p$. After some simple algebra one gets
$$\eqalign{
&\tr\vt_{m-2}\Big\vert_{X_l=F\delta_{l,m+1-k}} = \tr \Big(
{m-1\over 2}  (U_kF)'-U_{k+1}F +U_kF'\Big) \cr
&-\sum_{p=\max(0,k+1-m)}^{k+1} (-)^p
\left[ {m+1\over 2}\bin m-k+p-1 {m-k} -\bin m-k+p {m-k} \right]
\tr (FU_{k+1-p})^{(p)} \cr}
\eqn\qxvi$$
(where for $k=m$ one sets $U_{k+1}=U_{m+1}=0$). Note that terms like
$\tr (FU_{k-1-p})^{(p)} U_2$ cancelled against terms  $\tr U_2
(FU_{k-1-p})^{(p)}$ which would not have been the case without taking
the trace. Let first $k<m$ so that the sum over $p$ is from $0$ to
$k+1$. Separate the $p=k+1, p=0$ and $p=1$ terms from the sum (the
$p=k$ term vanishes since $U_1=0$). Using the identity
$$ {m+1\over 2}\bin m-k+p-1 {m-k} -\bin m-k+p {m-k} =
{m-1\over 2}\bin m-k+p-1 {p-1} -\bin m-k+p-1 p
\eqn\qxvii$$
one obtains
$$\eqalign{
&\tr\vt_{m-2}\Big\vert_{X_l=F\delta_{l,m+1-k}} = \tr \Big(
U_kF'+(k-1)  (U_kF)'+(-)^{k+1} {k-1\over 2}\bin m+1 {k+1} F^{(k+1)}\Big)
\cr
&+\sum_{p=2}^{k-1}(-)^p\left[ \bin m-k+p-1 p -{m-1\over 2}
\bin m-k+p-1 {p-1} \right] \tr (FU_{k+1-p})^{(p)} \ .\cr}
\eqn\qxviii$$
It is easily seen that for $k=m$ one also obtains the same equation
\qxviii. Now multiply by $\e$ and integrate to obtain $\gmd \{\int\e
T,\int \tr FU_k\}$ (cf. eq. \qx). Upon taking the functional derivative
with repect to $F$ and relabellng the summation index $p=k+1-l$ one
obtains \qxv.

Since the conformal properties \qxv\ are formally the same as in the
scalar case, and in the latter case it was possible to form
combinations $W_k$ that are spin-$k$ conformally primary fields, one
expects a similar result to hold in the matrix case. Indeed, one has the

\noindent
{\bf Theorem 4.6 :}   For matrices $A_1, A_2, \ldots A_r$ denote by
$S[A_1,A_2,\ldots,A_r]$ the completely symmetrized product normalized
to equal $A^r$ if $A_s=A$ for all $s=1,\ldots r$. Let
$$\eqalign{
W_k=&\sum_{l=2}^k B_{kl\, }U_l^{(k-l)} +\sum_{0\le p_1\le \ldots \le p_r
\atop \sum p_i+2r=k} (-)^{r-1} C_{p_1\ldots p_r}\,  S[U_2^{(p_1)},
\ldots, U_2^{(p_r)}]\cr
&+\sum_{0\le p_1\le \ldots \le p_r
\atop s\le l\le k-\sum p_i-2r} (-)^{r} D_{p_1\ldots p_r,l}\,
S[U_2^{(p_1)}, \ldots, U_2^{(p_r)},U_l^{(k-l-\sum p_i-2r)}]\cr }
\eqn\qxix$$
where the coefficiets $B_{kl}, C_{p_1\ldots p_r}$ and $D_{p_1\ldots
p_r,l}$ are the same as those given in ref. \DIZ\ for the scalar case, in
particular
$$B_{kl}=(-) ^{k-l}\ { \bin k-1 {k-l} \bin m-l {k-l} \over
\bin 2k-2 {k-l} }\ .
\eqn\qxx$$
Then the $W_k$ are spin-$k$ conformally primary
$n\times n$-matrix-valued fields, i.e.
$$\gmd \{\int \e T, W_k\}=-\e W_k' -k\e' W_k \ .
\eqn\qxxi$$
For $\s\in S^1$ one can define the modes
$(W_k)_s=\gmd\int_{-\pi}^\pi \rmd\s\, W_k(\s) e^{is\s}$ and the Virasoro
generators $L_r$ as  in \qv. Then one has equivalently
$$i\{L_r,(W_k)_s\}=\big( (k-1)r-s\big) (W_k)_{r+s}
\eqn\qxxia$$
where each $(W_k)_s$ is a $n\times n$-matrix.

\Proof   Note that in the scalar case eq. \qxix\ is identical to the
formula (2.11a) of ref. \DIZ\ if one identifies $U_l=-a_l,\, W_k=-w_k$.
The crucial property to prove the theorem is that \qxv\ is at most
linear in the $U_l$. It follows that
$$\eqalign{
&\gmd\{\int\e T, S\big[U_{k_1}^{(p_1)},U_{k_2}^{(p_2)}, \ldots ,
U_{k_r}^{(p_r)}\big]\}\cr
& =
\sum_{i=1}^r S\Big[ U_{k_1}^{(p_1)}, \ldots, U_{k_{i-1}}^{(p_{i-1})},
\gmd \{\int\e T, U_{k_i}\}^{(p_i)}, U_{k_{i+1}}^{(p_{i+1})}, \ldots ,
U_{k_r}^{(p_r)} \Big] \cr}
\eqn\qxxii$$
and since matrices commute under the symmetrization $S[\ldots ]$ one
may manipulate them just as in the scalar case. Hence eq. \qxix\ can be
proven exactly as in the scalar case and thus follows from the results
of ref. \DIZ.

\noindent
{\bf Examples :}   From the previous theorem and the results of ref.
\DIZ\ (Table I) one has explicitly:
$$\eqalign{
W_3&= U_3-{m-2\over 2} U_2' \cr
W_4&= U_4-{m-3\over 2} U_3'+{(m-2)(m-3)\over 10} U_2''
+{(5m+7)(m-2)(m-3)\over 10 m (m^2-1)} U_2^2 \cr
W_5&= U_5-{m-4\over 2} U_4'+{3(m-3)(m-4)\over 28} U_3''
-{(m-2)(m-3)(m-4)\over 84} U_2'''\cr
&+{(7m+13)(m-3)(m-4)\over 14 m (m^2-1)} (U_2W_3+W_3U_2) \ . \cr}
\eqn\qxxiii$$
Note that all coefficients are such that $W_k=0$ for $k>m$ if one sets
$U_l=0$ for $l>m$. These relations can be inverted to give
$$\eqalign{
U_3&= W_3+{m-2\over 2} U_2' \cr
U_4&= W_4+{m-3\over 2} W_3'+{3(m-2)(m-3)\over 20} U_2''
-{(5m+7)(m-2)(m-3)\over 10 m (m^2-1)} U_2^2 \cr
U_5&= W_5+{m-4\over 2} W_4'+{(m-3)(m-4)\over 7} W_3''
+{(m-2)(m-3)(m-4)\over 30} U_2'''\cr
&-{(7m+13)(m-3)(m-4)\over 14 m (m^2-1)} (U_2W_3+W_3U_2)
-{(5m+7)(m-2)(m-3)(m-4)\over 20 m (m^2-1)}(U_2^2)' \ . \cr}
\eqn\qxxiv$$

\chapter{{The Poisson bracket algebra of $U_2$ and $W_3$ for arbitrary
$m$}}

{}From the previous subsection one might have gotten the impression that
the matrix case is not very different from the scalar case. This is
however not true. In the previous subsection only the conformal
properties, i.e. the Poisson brackets with $T={1\over 2} \tr \id U_2$
were studied, and since the unit-matrix $\id$ always commutes, most of
the new features due to the non-commutativity of matrices were not
seen. Technically speaking, only $\tr \vt$ was needed, not $\vt$
itself. In this section, I will give the Poisson brackets, for the (more
interesting) reduction to $U_1=0$, of any two matrix elements of $U_2$
or $U_3$, or equivalently of $U_2$ or $W_3$, for arbitrary $m$. In the
case $m=3$ this is the complete algebra, giving a matrix generalization
of Zamolodchikov's $W_3$-algebra.

The Poisson bracket algebra will again be obtained from \dxxx. Since
$\{\int\tr FU_2,\int\tr GU_2\}$ was already computed in the previous
section, eq. \qii, here I will need to compute
$\{\int\tr FU_3,\int\tr GU_2\}$ and
$\{\int\tr FU_3,\int\tr GU_3\}$ only. Thus all one needs is $\vt_j$ for
$j=m-2,\, m-3$ and with $X_l=F\delta_{l,m-2}$. For $j=m-2$ all relevant
coefficients $C_{m-q-p,m-2}^{m-q,m-2}$ are given in Lemma 4.4 (with
$k=3$), and the computation of $\vt_{m-2}\vert_{X_l=F\delta_{l,m-2}}$
proceeds as in the proof of Proposition 4.5 (but without discarding
terms that vanish upon taking the trace. It is then straightforward to
obtain $\{\int\tr FU_3,\int\tr GU_2\} =
a \int\tr \vt_{m-2}\vert_{X_l=F\delta_{l,m-2}} G$, and using the
antisymmetry of the bracket also
$$\eqalign{
\{\int\tr FU_2,&\int\tr GU_3\} = a \int\tr\Big(
-{1\over m} [F,U_2]\d^{-1}[G,U_3] -{m-2\over m} F[G,U_2]U_2 \cr
&-[F,G]U_4 +{(m-1)(m-2)\over 6}[F,G'']U_2-{m-1\over 2}[F',G]U_3\cr
&+(2F'G-G'F)U_3+(m-2)F''GU_2-\bin m+1 4 G F^{(4)} \Big) \ . \cr}
\eqn\ci$$
Of course, if $F={1\over 2}\e\, \id$ all commutator terms vanish and upon
taking the functional derivative with respect to $G$ one recovers eq.
\qxv\ for $k=3$ (recall $a=-2\gd$).

Consider now $\vt_j$ for $j=m-3$ and with $X_l=F\delta_{l,m-2}$. The
relevant sum in \dxxx\ containing the $C$-coefficients is
$$\sum_{p=0}^5\ \sum_{q=\max(0,p+m-5)}^{\min(m,p+2m-5)}
C_{q-p,m-2}^{q,m-3} U_{m-q} (FU_{q-(p+m-5)})^{(p)} \ .
\eqn\cii$$
For $m\ge 5$ the sum over $q$ is simply $\sum_{q=p+m-5}^m$. For $m<5$,
one can still formally write $\sum_{q=p+m-5}^m$ if one defines $U_k=0$
for $k<0$ or $k>m$. In any case one has $q-p\ge m-5$. From the
definition \dxiv\ of $S_{q-p,m-2}^{q,m-3}$ one sees that it vanishes
for $q-p>m-3$, and that it is a sum of at most three terms for $m-3\ge
q-p\ge m-5$. Thus all relevant $S_{q-p,m-2}^{q,m-3}$ can be easily
computed directly to obtain the $C_{q-p,m-2}^{q,m-3}$ appearing in
\cii. I will not give all the details here. The result is
$$\eqalign{
\{\int\tr FU_3,\int\tr GU_3\} = a \int\tr\Big( &
-{1\over m} [F,U_3]\d^{-1}[G,U_3] +{2\over m}(FU_2GU_3-FU_3GU_2)\cr
&-{m-2\over m} [F,G]U_2U_3 +{2(m-2)\over m} (FU_2)' GU_2\cr
&-[F,G]U_5+2(F'G-G'F)U_4 +(G''F-F''G)U_3\cr
&+{(m-1)(m-2)\over 6}\left([F'',G]+[F,G'']\right) U_3  \cr
&-{(m-1)(m-2)\over 3}(F'''G-G'''F)U_2 \cr
&+{4m^2-3m-7\over 30} \bin m 3  G F^{(5)} \Big) \cr}
\eqn\ciii$$
where $U_5=0$ for $m=4$ and $U_5=U_4=0$ for $m=3$. Note the obvious
antisymmetry under exchange of $F$ and $G$.

One may also reexpress the Poisson brackets \qii, \ci\ and \ciii\ using
the primary fields $W_k$. Substituting eqs. \qxxiv\ it is
straightforward algebra\foot{One uses the obvious relations
$\{f,\int\tr GU_2'\}=-\{f,\int\tr G'U_2\}$ and also the fact that $\int
A\d^{-1}B=-\int (\d^{-1}A) B$.}
to obtain
$$\eqalign{
\{\int\tr FU_2,\int\tr GU_2\}= &a\int\tr \Big(
-{1\over m} [F,U_2]\d^{-1}[G,U_2]-[F,G]W_3\cr
&+{1\over 2}(F'G-G'F+GF'-FG')U_2 -{1\over 2} \bin m+1 3  G F''' \Big)
\cr}
\eqn\civ$$
$$\eqalign{
\{\int\tr FU_2,\int\tr GW_3\} = &a \int\tr\Big(
-{1\over m} [F,U_2]\d^{-1}[G,W_3] -{4(m^2-4)\over 5 m(m^2-1)}
[F,G]U_2^2\cr
& -[F,G]W_4 +(F'G+GF'-{1\over 2}FG'-{1\over 2}G'F)W_3\cr
&+{m^2-4\over 5}\Big( -{1\over 4}[F',G']+{1\over 12}[F,G'']
+{1\over 2}[F'',G]\Big) U_2 \Big)  \cr}
\eqn\cv$$
$$\eqalign{
\{\int&\tr FW_3,\int\tr GW_3\} =a \int\tr \Big(
-{1\over m}[F,W_3]\d^{-1}[G,W_3] \cr
&+{2\over m}(FU_2GW_3-GU_2FW_3)
-{11m^2-71\over 7m(m^2-1)}[F,G](W_3U_2+U_2W_3)\cr
&+{m^2-4\over 4m}(F'U_2GU_2-G'U_2FU_2)
+{(m-2)^2\over 4m}\left( [F,G]U_2'U_2 +(FG'-GF')U_2^2 \right)\cr
&+{(5m+7)(m-2)(m-3)\over 10 m (m^2-1)}(FG'-GF'+G'F-F'G)U_2^2\cr
&-[F,G]W_5 - (FG'-GF'+G'F-F'G) W_4
+{m^2-16\over 84}\left( 2[F,G]''-7[F',G']\right) W_3\cr
&+{m^2-4\over 120}\big( 2(FG'''+G'''F-F'''G-GF''')
+3(F''G'+G'F''-F'G''-G''F')\big) U_2\cr
&+{1\over 6} \bin m+2 5  GF^{(5)} \Big)\ . \cr}
\eqn\cvi$$
Note again, as a consistency check, that \civ\ and \cvi\ are
obviously antisymmetric under $F\leftrightarrow G$. Of course, one has
to set $W_5=0$ for $m=4$, and $W_5=W_4=0$ for $m=3$. For $m=3$,
equations \civ-\cvi\ are the complete algebra and reproduce the equations
\uiii-\uv\ written in the introduction. Remark that for $F={1\over 2}
\e\id$, the r.h.s. of eq. \cv\ reduces to  $-\gd\int\tr (2\e' G-\e
G')W_3$ (recall $a=-2\gd$), confirming once again that every matrix
element of $W_3$ is a conformal primary field of dimension 3. If both
$F$ and $G$ are proportional to the unit matrix $F=f\id,\, G=g\id$, with
scalar $f,g$, most of the terms in \cvi\ disappear and one has
$$\eqalign{
\{\int f\, \tr W_3,\int g\, \tr W_3\}
=a \int\tr \Big(
&(m^2-4)(f'g-g'f)\left( {8\over 5m(m^2-1)}\tr U_2^2-{1\over
15} T''\right)\cr
&+2(f'g-g'f)\tr W_4 +{m^2-4\over 6}(f''g'-g''f')T\cr
&+{n\over 6} \bin m+2 5  gf^{(5)} \Big) \cr}
\eqn\cvii$$
where I used $\tr\id=n$ and $\tr U_2=2T$. This looks similar to the
corresponding bracket in the scalar case ($n=1$). It is different,
however, since $\tr U_2^2\ne (\tr U_2)^2$ in the matrix case. Thus the
standard (scalar) $W$-algebra is {\it not} a subalgebra of the $n\ne
1$-algebras.\foot{There is one trivial exception: for $m=2$ one only
has $\{\tr FU_2, \int \tr GU_2\}$ which is linear, and hence $T={1\over
2}\tr U_2$ forms a closed subalgebra, the Virasoro algebra discussed in
section 4.}

\chapter{{Other algabras, restrictions and concluding remarks}}

\section{Hermiticity of the $W_k$}

The (quantum) Virasoro algebra $[L_r,L_s]=(r-s)L_{r+s}+{c\over
12}(r^3-r)\delta_{r+s,0}$ is compatible with the hermiticity condition
$L_r^+=L_{-r}$ (where the hermitian conjugation refers to the inner
product on the Hilbert space on which the Virasoro generators act).
Similarly, the Poisson bracket version \qvii\ is compatible with the
reality condition $L_r^*=L_{-r}$. For real $\gd$, i.e. real scale
factor $a$ this is equivalent to $T^*=T$. The natural extension of this
condition to the matrix case is the hermiticity condition $U_2^+=U_2$
(where now hermitian conjugation is simply the hermitian conjugation of
the $n\times n$-matrix). Assuming the matrix $U_2$ to be hermitian is
also natural when studying (for $m=2$) the resolvent of $L=-\d^2+U_2$
[\MCK]. More generally one has the

\noindent
{\bf Conjecture 6.1 :}
For real scale factor $a$, the second \GD bracket, eq. \dxxx, is
compatible with the hermiticity conditions
$$ U_2^+=U_2 \quad , \quad W_k^+=(-)^k\,  W_k\ ,\ \ k\ge 3 \ .
\eqn\si$$

\Lemma 6.2   A sufficient condition for the bracket
$$\{\int\tr FW_k, \int\tr GW_l\}=\int\tr P_{kl}(F,G,W_r)
\eqn\sii$$
to be compatible with the conditions \si\ is
$$\int\tr P_{kl}(F,G,W_r)^+
=(-)^{k+l}\int\tr P_{kl}(F^+,G^+,(-)^rW_r^+) \ .
\eqn\siii$$

\Proof   Compatibility means\foot{Recall the example of the Virasoro
algebra where complex conjugation using $L_r^*=L_{-r}$ gives the same
algebra upon relabelling $r\to -r, s\to -s$ which corresponds to
replacing $f=e^{ir\s}\to f^*=e^{-ir\s}$ and $g=e^{is\s}\to
g^*=e^{-is\s}$.} that if one takes the hermitian conjugate
of \sii\ and uses \si\ one gets back the {\it same} bracket \sii\ with
the same functional $P_{kl}$ and $F$ and $G$ replaced by $F^+$ and
$G^+$.
Taking the hermitian conjugate of \sii\ yields, using
\si\ and \siii,
$$\eqalign{
(-)^{k+l}\{\int\tr F^+W_k, \int\tr G^+W_l\}
&=\int\tr P_{kl}(F,G,W_r)^+ \cr
=(-)^{k+l}\int\tr P_{kl}(F^+,G^+,(-)^r W_r^+)
&=(-)^{k+l}\int\tr P_{kl}(F^+,G^+,W_r) \cr}
\eqn\siv$$
which is again eq. \sii\ with $F$ and $G$ replaced by $F^+$ and
$G^+$. Thus \siii\ is a sufficient condition.

\Lemma 6.3   All the second \GD brackets of the $U_2$ or the $W_k$ given
explicitly in this paper, namely \qxxi\ and eqs. \civ\ to \cvi, are
compatible with the hermiticity conditions \si.

\Proof   By the previous Lemma one has to check whether \siii\ is
satisfied. For eq. \qxxi\ this is trivial. In general however, the
condition \siii\ is non-trivial. Consider e.g. eq. \cvi. Since \siii\
is a linear condition it can be checked on groups of terms separately.
For $[F,G]W_5$ one has e.g.
$\tr\left([F,G]W_5\right)^+
=\tr\left([F,G]^+W_5^+\right)
=-\tr\left([F^+,G^+]W_5^+\right)
=\tr\left( [F^+,G^+](-)^5W_5^+\right)
=(-)^{3+3}\tr\left( [F^+,G^+](-)^5W_5^+\right)$
while for $(F'G+GF')W_4$ one has
$\tr \big((F'G+GF')W_4\big)^+ =
(-)^{3+3}\tr\left( ({F^+}'G^+ +G^+{F^+}')(-)^4W_4^+\right)$
and the condition \siii\ is satisfied.
On the other hand, a term like $\tr [F,G]
W_4'$ would lead to the wrong sign, and indeed it does not appear
(although it has the correct antisymmetry properties under
$F\leftrightarrow G$ and the correct ``naive" dimension). One can
easily check that all terms on the r.h.s. of \civ-\cvi\ have the
required properties. The only slightly non-trivial terms in \cvi\ are
$\int\tr\left([F,G]U_2'U_2 + (FG'-GF')U_2^2\right)$. Here one needs
to integrate by parts to show that \siii\ is satisfied.

\section{Restrictions and other algebras}

In section 2, I already discussed the restriction (reduction) to
$U_1=0$, and in section 3 the corresponding reduction $\sum_{i=1}^m P_i
=0$. The original $V$-algebra \ui\ corresponds to $m=2,\ n=2,\ U_1=0$
and furthermore  $\tr\s_3U_2=0$. Recall that the reduction to $U_1=0$
was implemented by determining $X_m$ (which formerly was $\dd f {U_1}$)
such that
$\{U_1, f\}\vert_{U_1=0}$ vanishes. Similarly if one
decomposes the $2\times 2$-matrices
$$U_2\equiv U=\pmatrix{ T+V_3& -\rd
\vp\cr -\rd\vm& T-V_3\cr} \quad , \quad
\dd f U =\pmatrix{
(F_0+F_3)/2&-F^-/\rd \cr -F^+/\rd & (F_0-F_3)/2\cr} \ ,
\eqn\sv$$
the reduction to
$V_3=0$ is achived by determining $F_3$ (which formerly was $\sim \dd f
V_3$) such that $\{V_3,f\}\vert_{V_3=0}=0$. It can be seen from \dxxxiv\
with $n=2$ or from \dxxxvi\ that this implies $F_3=0$. (This contrasts
with the reduction to $U_1=0$ where $X_m$ was a non-trivial function of
the $X_1,\ldots X_{m-1}$.) Thus to obtain the reduction to $\tr
\s_3U=2V_3=0$ it is enough to simply set $V_3=F_3=0$ in eq. \dxxxiv\ or
\dxxxvi. The result is equivalent to the original $V$-algebra \ui.

Other reductions can be achieved by not only taking $U_1=0$, but by
reducing to the symplectic submanifold where several $U_k$ vanish, e.g.
$U_k=0$ for all odd $k$. Another, and probably more fruitful
approach is to take advantage of the Miura transformation and impose
conditions on the $P_i$. Here, I studied the $A_{m-1}$-type reduction
$\sum_{i=1}^mP_i=0$. But one can study other reductions, like
$P_{m+1-i}=-P_i$ that correspond to the Lie algebras $B_{{m-1\over 2}}$
if $m$ is odd, and to $C_{{m\over 2}}$ if $m$ is even. They should lead
to matrix generalizations of the $WB_{{m-1\over 2}}$ and
$WC_{{m\over 2}}$-algebras.

\section{Concluding remarks}

In this paper, I have given matrix generalizations of the
well-known $W_m$-algebras by constructing the second \GD bracket
associated with a matrix linear differential operator of order $m$.
Upon reducing to $U_1=0$, the non-commutativity of matrices implies the
presence of non-local terms in the algebra.

One always has a Virasoro subalgebra generated by $T={1\over 2}\tr
U_2$, and all other (``orthogonal") combinations of matrix elements of
$U_2$ (i.e. $\tr A_\mu U_2$ with $\tr A_\mu =0$) are spin-two
conformally primary fields, while the $U_k,\, k\ge 3$ can be combined
into matrices $W_k,\, k\ge 3$ which are (matrices of) spin-$k$ conformal
primary fields. I have given the complete Poisson bracket algebra for
$m=3$, and,  for all $m$, the Poisson brackets involving $U_2$, $W_3$.

A Miura transformation relates these Poisson brackets of the $U_k$ to
much simpler ones of a set of $n\times n$-matrices $P_i$. Contrary to
the case $n=1$, the $P_i$ are not free fields. However, for $m=n=2$,
$U_1=0$ and $\tr\s_3U_2=0$ a simple free-field realization for
$P_1=-P_2$ was given in refs. \NAT\ and \MCK\ in terms of vertex
operator-like fields. In the general case, it is not clear how to give
such a free-field realization. The main difficulty is to realize the
non-local terms. Comparing the more general case \dxxxvi\ with \ui\ one
sees the origin of the difficulty: $\{V^+(\s),V^-(\s')\}\sim \es
V^+(\s)V^-(\s')+\ldots$ can be realized by $V^\pm(\s)\sim e^{\mp i\rd
\varphi(\s)}\, R(\d\varphi)$ where $R(\d\varphi)$ is some
differential polynomial in $\d\varphi$ and $\varphi$ is a free field. On
the other hand, $\{V_3(\s),V^\pm(\s')\}\sim \es
V^\pm(\s)V_3(\s')+\ldots$ cannot be realized by this type of vertex
operator construction, since the arguments $\s$ and $\s'$ of $V^\pm$
and $V_3$ have been exchanged. This kind of relation is however very
reminiscent of the braiding relations of chiral screened vertex
operators in conformal quantum field theories
\REF\FFK{G. Felder, J. Froehlich and G. Keller, {\it Braid matrices
and structure constants for minimal conformal models}, Commun. Math.
Phys. {\bf 124} (1989) 647.}
\REF\BILBR{A. Bilal, {\it Fusion and
braiding in extended conformal theories}, Nucl. Phys. {\bf B330} (1990)
399; {\it Fusion and braiding  in extended conformal theories (II):
Generalization to chiral screened vertex operators labelled by arbitrary
Young tableaux} Int. J. Mod. Phys. {\bf A5} (1990) 1881.}
[\FFK,\BILBR] and it might well be that a free field realization,
involving screening type integrals, can be given. Once a free field
realization is found, one can try to quantize the structures described
in this paper. This will certainly lead to most interesting developments.

\Appendix{A}

In this appendix, I recall some well-known properties of
pseudo-differential operators and adapt them to the matrix case.

\Lemma A.1   Let $A,B$ be some matrix-valued pseudo-differential
operators, i.e. let $A=\sum_{i=-\infty}^l a_i\d^i,
B=\sum_{j=-\infty}^k b_j\d^j$ with $a_i, b_j$ some matrix-valued
functions. Then there exists a matrix-valued function $h$ such that
$$\tr\res [A,B]=(\d h)\equiv h' \ .
\eqn\ai$$
In particular, if the integral of a total derivative vanishes one has
$\int\tr\res A B =\int\tr\res B A$.

\Proof    The proof is a straightforward
generalization of the scalar case (see e.g. p.9 of
\REF\DIKBOOK{L.A. Dickey, {\it Soliton equations and hamiltonian
systems}, World Scientific, 1991.}
ref. \DIKBOOK). By linearity, it is sufficient to prove this for
monomials $A=a_i\d^i, B=b_j\d^j$. If $i,j\ge 0$ or $i+j<1$ the residue
obviously vanishes and $h=0$. Hence let $i\ge 0$ and $j<0$ with $i+j\ge
1$. Then using eq. \dii:
$$\eqalign{
\res[A,B]&=\res (a_i\d^ib_j\d^j-b_j\d^ja_i\d^i)\cr
&=\res\left(\sum_{p=0}^i\bin i p a_i b_j^{(i-p)}\d^{j+p}-
\sum_{s=0}^\infty (-)^s \bin -j+s-1 s b_j a_i^{(s)}\d^{j-s+i}\right)\cr
&=\bin i {i+j+1} \left(a_ib_j^{(i+j+1)}+(-)^{i+j}b_j a_i^{(i+j+1)}
\right) \ . \cr}
\eqn\aii$$
Let
$$h=\bin i {i+j+1} \sum_{p=0}^{i+j} (-)^p a_i^{(p)} b_j^{(i+j-p)} \ .
\eqn\aiii$$
Upon taking $h'$ one sees that there are cancellations between two
successive $p$  terms, and only part of the terms with $p=0$ and $p=i+j$
survive. The result equals the r.h.s. of \aii, except for the order of
the matrices. Upon taking the trace, eq. \ai\ follows.

\Lemma A.2   Let $A=\sum_{i=-\infty}^l a_i\d^i$. One can always rewrite
$A=\sum_{i=-\infty}^l \d^i \tilde a_i$. One has $\tilde a_{-1}=
a_{-1}$.

\Proof   Decompose  $A=\sum_{i=0}^l \d^i \tilde a_i +\d^{-1}\tilde
a_{-1} + \sum_{i=-\infty}^{-2} \d^i \tilde a_i$ and use \dii\ to see
that $\tilde a_{-1} \d^{-1}$ is the only term containing $\d^{-1}$.

\Lemma A.3   If $h$ is a matrix-valued function and $A$ a matrix-valued
pseudo-differential operator then
$$\eqalign{
\res (Ah)&= (\res A) h\cr
\res (h A)&= h\, \res A\ .\cr}
\eqn\aiv$$

\Proof   The first identity is shown as for Lemma A.2 and the second is
obvious.

\Lemma A.4   Let $h$ and $A$ be as before. Then
$$\res A=(\d A_-)_+  =(A_-\d)_+= ((\d-h)A_-)_+= (A_-(\d-h))_+ \ .
\eqn\av$$

\Proof   Writing $A_-=a_{-1}\d^{-1}+A_{--}$ where
$A_{--}=\sum_{i=-\infty}^{-2}a_i\d^i$ the proof is obvious.

\Lemma A.5   Let $h$ and $A$ be as before. Then
$$[\d-h, \res A]=\res [\d-h,A] \ .
\eqn\avi$$

\Proof   Let $\dh=\d-h$. Then from the previous Lemma, and since
$A_-=A-A_+$
$$\eqalign{
[\d-h,\res A]&=[\dh, \res A]=\dh (A_-\dh)_+ -(\dh A_-)_+\dh\cr
&=\dh(A\dh)_+ -\dh(A_+\dh)_+ -(\dh A)_+\dh +(\dh A_+)_+\dh\cr
&=\dh(A\dh)_+ -\dh A_+\dh -(\dh A)_+\dh +\dh A_+\dh\cr
&=\dh(A\dh)_+ -(\dh A)_+\dh \cr}
\eqn\avii$$
while
$$\eqalign{
\res[\d-h, A]&=\res \dh A - \res A\dh =\left((\dh A)_-\dh\right)_+
-\left(\dh(A\dh )_-\right)_+  \cr
&=\left(\dh A\dh\right)_+ -\left((\dh A)_+\dh\right)_+
-\left(\dh A\dh  \right)_+ +\left(\dh(A\dh )_+\right)_+  \cr
&=-(\dh A)_+\dh + \dh (A\dh)_+ \ .\cr}
\eqn\aviii$$
Comparing \avii\ with \aviii\ completes the proof.

\Appendix{B}

In this appendix, I prove Lemma 2.4 and some other useful formulas for
the $S_{r,l}^{q,j}$. In both cases considered in Lemma 2.4 one has
$r\ge 0$ and $q\le j$ so that
$$S_{r,l}^{q,j}=\sum_{s=r}^q (-)^{s-r}\bin s-r+l-1 {l-1} \bin q s
=\sum_{s=0}^{q-r} (-)^s \bin s+l-1 {l-1} \bin q {s+r}  \ .
\eqn\bi$$
Note that for $q<r$, $S_{r,l}^{q,j}$ was defined to vanish, as do the
r.h.s. of \dxx\ and \dxxi. So let's suppose $q-r\ge 0$. Then the r.h.s.
of \bi\ is
$$\sum_{s=0}^{q-r}(-)^s {(s+l-1)!\over (l-1)! s!}{q!\over
(s+r)!(q-r-s)!}
={q!\over (l-1)!(q-r)!} \sum_{s=0}^{q-r}(-)^s {(s+l-1)!\over (s+r)!}
\bin q-r s  \ .
\eqn\bii$$

Let first $l>r$, so that ${(s+l-1)!\over (s+r)!}=(s+l-1)(s+l-2)\ldots
(s+r+1)$ and
$$\eqalign{
\sum_{s=0}^{q-r}(-)^s {(s+l-1)!\over (s+r)!} \bin q-r s
&=\left( {d\over d x}\right)^{l-1-r}\
\sum_{s=0}^{q-r} x^{s+l-1} (-)^s  \bin q-r s  \Big\vert_{x=1}\cr
&=\left( {d\over d x}\right)^{l-1-r} x^{l-1}(1-x)^{q-r}\Big\vert_{x=1} \
.\cr}
\eqn\biii$$
This vanishes if $l-1<q$ while for $l-1\ge q$ it equals $(-)^{q-r}
{(l-1)!(l-1-r)!\over q! (l-1-q)!}$. Inserting this result into \bii\
one has
$$S_{r,l}^{q,j}=(-)^{q-r} \bin l-1-r {q-r}
\eqn\biv$$
which proves \dxxi.

Let now $1\le l\le r$, so that
${(s+l-1)!\over (s+r)!}={1\over (s+r)(s+r-1)\ldots (s+l)}$ and
$$\eqalign{
\sum_{s=0}^{q-r}(-)^s {(s+l-1)!\over (s+r)!} \bin q-r s
&=\left( \d_x^{-1}\right)^{r-l+1}
\sum_{s=0}^{q-r} (-)^s x^{s+l-1}   \bin q-r s  \Big\vert_{x=1}\cr
&=\left( \d_x^{-1}\right)^{r-l+1} x^{l-1}(1-x)^{q-r}\Big\vert_{x=1}
\cr}
\eqn\bv$$
where here $\left( \d_x^{-1}\right)^p h(x) =\int_0^x\rmd x_1
\int_0^{x_1} \rmd x_2 \ldots \int_0^{x_{p-1}}\rmd x_p h(x_p)$. For non
negative integers $a,b$ one has
$$\int_0^1\rmd x_1
\int_0^{x_1} \rmd x_2 \ldots \int_0^{x_{p-1}}\rmd x_p\,
x_p^a (1-x_p)^b ={a! (b+p-1)!\over (a+b+p)! (p-1)!}
\eqn\bvi$$
so that expression \bv\ equals ${(l-1)!(q-l)!\over q! (r-l)!}$.
Inserting this into the r.h.s. of \bii\ yields $\bin q-l {q-r}$ which
proves \dxx.

Another immediate consequence of the definition \dxiv\ of
$S_{r,l}^{q,j}$ is, for ${p\ge 0}$ and ${l\ge 1}$
$$\eqalign{
S_{m-p,l}^{m,m-1}&=\sum_{s=\max(0,m-p)}^{m-1}(\ldots)=
\sum_{s=\max(0,m-p)}^{m}(\ldots) - (\ldots)\vert_{s=m}\cr
&=S_{m-p,l}^{m,m} -(-)^p \bin p+l-1 {l-1} \ . \cr}
\eqn\bvii$$
Furthermore, for $p\ge 1, l\ge 1$ one has
$$\eqalign{
S_{m-p,l}^{m,m-2}&=\sum_{s=\max(0,m-p)}^{m-2}(\ldots)=
\sum_{s=\max(0,m-p)}^{m}(\ldots) - (\ldots)\vert_{s=m}
- (\ldots)\vert_{s=m-1} \cr
&=S_{m-p,l}^{m,m} -(-)^p \bin p+l-1 {l-1}+(-)^p\, m\bin p+l-2 {l-1} \ .
\cr}
\eqn\bviii$$
If now $p=0$, one has $S_{m,l}^{m,m}=1$, $S_{m,l}^{m,m-2}=0$ and $\bin
l-2 {l-1} =0$, so that eq. \bviii\ remains true also for $p=0$.

\noindent
{\bf Note Added}

After submission of this paper I became aware of the work by Feher
et al
\REF\FEH{L. Feh\'er, J. Hanad and I. Marshall, {\it Generalized
Drinfeld-Sokolov reductions and KdV type hierarchies}, Commun. Math.
Phys. {\bf 154} (1993) 181.}
[\FEH]. These authors derive my starting equation \dv\ from a
generalized Drinfeld-Sokolov reduction of a $nm\times nm$-matrix
first order differential operator. They also mention relations
to $W$-algebras and give explicit (KdV-flow) formulas for $m=2$.
They do not, however, address my main point of interest here, namely
the reduction to $U_1=0$.
As already repeatedly
emphasized, it is the reduction to $U_1=0$ that brings about the new
non-local terms, characteristic for the $V$-algebras.

\refout
\end